\documentclass[aps, twocolumn, amsmath,superscriptaddress]{revtex4}
\usepackage{dcolumn}
\usepackage{bm}
\usepackage{graphicx}
\usepackage{color}
\usepackage{amsfonts}
\usepackage{booktabs}
\usepackage{boldline}
\DeclareGraphicsExtensions{. jpg,. pdf, . mps, . png, . eps, . ps, . EPS}

\begin{document}
\def\be{\begin{equation}}
\def\ee{\end{equation}}

\def\bc{\begin{center}} 
\def\ec{\end{center}}
\def\bea{\begin{eqnarray}}
\def\eea{\end{eqnarray}}
\newcommand{\avg}[1]{\langle{#1}\rangle}
\newcommand{\Avg}[1]{\left\langle{#1}\right\rangle}

\newcommand{\ra}[1]{\renewcommand{\arraystretch}{#1}}

\newcommand{\NoteFR}[1]{{\color{red}\tt (FR: {#1})}}

\newcommand{\gin}[1]{{\color{magenta}\tt (GIN: {#1})}}

\title{Controlling the uncertain response of real multiplex networks to random damage}

\author{Francesco Coghi}
\affiliation{School of Mathematical Sciences, Queen Mary University of London, London E1 4NS, UK}

\author{Filippo Radicchi}
\affiliation{Center for Complex Networks and Systems Research, School
  of Informatics, Computing, and Engineering, Indiana University, Bloomington, IN 47408, USA}

\author{Ginestra Bianconi}
\affiliation{School of Mathematical Sciences, Queen Mary University of London, London E1 4NS, UK}

\begin{abstract}
We reveal large fluctuations in the response of real multiplex networks to random damage of nodes.  These results indicate that the average response to random damage,  traditionally considered
in mean-field approaches to percolation, is a  poor metric of 
system robustness.  We show instead that a large-deviation approach to percolation 
provides a more accurate characterization of system robustness. 
We identify an effective percolation threshold at which we observe a
clear abrupt transition separating two distinct regimes in which the most likely response 
to damage is either a functional or a dismantled multiplex network. We leverage our findings to 
propose a new metric, named safeguard centrality, able to 
single out the nodes that control the response of the entire multiplex
 network to random damage. We show that  safeguarding the function of
 top-scoring nodes is sufficient to prevent system collapse.
\end{abstract}

\maketitle
\section{Introduction}

Civil infrastructures, transportation networks, financial networks, as
well as cell molecular networks and  brain networks,
are all good examples of multiplex networks, i.e., complex systems whose topology 
can be meaningfully represented as a composition of many 
interacting network layers~\cite{Bianconi_book,PhysRep,Kivela,Goh_review,PRE}.
A central topic in the study of multiplex networks is 
the characterization of their robustness \cite{Havlin}. This problem is usually approached 
with percolation theory, where the
macroscopic connectedness of the system 
is studied as a 
function of the microscopic damage of 
system elements. The simplest scenario considered 
in percolation studies of multiplex networks assumes that nodes are 
initially damaged with probability $f$ (alternatively, one may assumes that nodes
are not damaged with probability $p = 1 -f$).
Depending on the topology of the system and the value of the
probability $f$, the initial damage of nodes may trigger further 
damaging avalanches in the system, eventually 
leading to the complete failure of the multiplex \cite{Havlin}.
On networks with infinite size, it 
has been shown that percolation yields a 
discontinuous hybrid 
transition  {of the Mutually Connected Giant Component (MCGC), }thus radically different
from the usual continuous transition observed
in isolated networks~\cite{Havlin,Dorogovtsev,Goh,HavlinEPL,Havlin2,Son,BD1,BD2,Baxter2016,Cellai2016,RadicchiBianconi,BianconiRadicchi,Cellai2013,Radicchi2015,Goh_comment,Makse_NP}. The discontinuity of the transition indicates that 
multiplex networks are significantly more fragile 
than their single layers taken in isolation. 
The reason is that, at the 
percolation transition, a multiplex network is affected 
by large avalanches of failure that 
suddenly dismantle the whole network leading to a discontinuous phase transition~\cite{Havlin}.
This result is central in the study of percolation and is playing a major role in the active research field aiming at identifying  dynamical rules that can change the nature of  phase transitions from continuous to discontinuous \cite{Souza,Arenas,RadicchiArenas,Vito,Doro_kcore,Rizzo}. 

Percolation theory, on single-layer as well as 
on multiplex networks, is traditionally 
studied in the {\em mean-field} approach by 
characterizing the average response of a network to initial damage \cite{crit,Lenka}.
This approach is totally justified in the infinite network limit where percolation 
is self-averaging, i.e., the fluctuations around the mean behaviour are vanishing.
However, the interest in the percolation transition is often driven by applications  
which always involve finite (and sometime not too large) networks \cite{Havlin,Radicchi2015,RadicchiBianconi,Wei}.
Further in practical applications, the prediction of eventual, 
even if extremely  rare, 
catastrophic failures is way more important than the 
characterization of the average behavior of a system.

To provide a pragmatic characterization of the response to damage 
of real networks, recent papers, such as Refs.~\cite{Fluct,Bianconi2018,Makse,Dismantling},
on percolation in single-layer networks 
went beyond the standard  {\em mean-field} approach.
In Ref.~\cite{Bianconi2018},
a  theoretical  framework based on large deviation theory 
was proposed to predict the probability distribution $\pi(R)$ 
for the relative size $R$ of the giant component in single instances 
of the percolation model
on a given network. The approach allows for the theoretical computation of $\pi(R)$ 
starting from any real network datasets.  {Results of the paper show that  the average value of
$\pi(R)$, hence the main observable of the mean-field
approach, may not be the best metric to study system robustness.}
Also, optimal percolation defines 
a problem that goes beyond the traditional 
percolation model~\cite{Makse,Dismantling}. Optimal percolation 
refers to the identification of 
the {\em optimal (minimal) structural node set} whose removal
leads a destruction of the entire network. In this sense, optimal 
percolation is
the problem of identifying the one rare realization of damage  
that has the most dramatic consequences for the network.  

In the context of multiplex networks several work 
have started to characterize the response to damage  beyond the
mean-field approach.  {
In Ref.~\cite{Kahng}, the authors studied finite-size effects in
multiplex network 
percolation focusing both on the 
final size of the MCGC and the avalanche distribution.}
In Ref. \cite{Krapivsky}, Kitsak and collaborators
analyzed the stability of the 
MCGCs  by considering the overlap among a large 
number of MCGCs resulting from  initial 
damage configurations drawn from the same distribution.
Optimal percolation was recently extended to multiplex 
networks in Ref.~\cite{Radicchi2017}.   The main finding is that 
optimal percolation on a multiplex network is a
rather distinct problem from the one defined 
for the individual network layers that compose the multiplex. 
Further work in this direction was
presented in  Ref.~\cite{Mendes2018}.

In this paper, we aim at providing 
a novel characterization of the percolation transition 
in multiplex networks. The approach we propose 
is similar to one already used in
Ref.~\cite{Bianconi2018} for isolated networks, thus placing
emphasis on  
large deviation properties of percolation. We consider a process
where  a fraction $f=1-p$ of nodes is initially damaged, and 
show that the probability $\pi(R)$ that the relative
size of the MCGC equals $R$ is
bimodal. 
The two peaks of the distribution $\pi(R)$ correspond to the
percolating and non-percolating phases, and in specific ranges 
for the parameter $p$ they quantify an equal likelihood for the system to be in the
functioning or nonfunctioning regimes. In this respect, 
the mean-field percolation diagram where the average value
$\bar{R}$ is plotted against $p$ provides distorted information about the 
robustness of the system, making
it look less fragile than actually is.  {An alternative phase diagram 
can be instead created by replacing $\bar{R}$ with $\hat{R}$, i.e., the mode of $\pi(R)$.}
In the phase diagram, $\hat{R}$ displays 
a clear discontinuity.  {However, the large fluctuations observed in single instance of percolation cannot be captured entirely by this single metric. 
The actual robustness properties of
 the network can be fully understood only by looking at the distribution
 $\pi(R)$. Nevertheless, as a matter of fact, the combination 
of the two metrics $\bar{R}$ with $\hat{R}$ results in an effective tool for 
the characterization of its robustness profile. 
These numbers respectively underestimate and overestimate the fragility of
the system, allowing for a direct interpretation 
the fragility properties of the network in practical contexts. In
this respect, } we identify an  
{\em effective critical point} $p_c$ with the discontinuity 
of $\hat{R}$, and show that, for $p=p_c$, the system is 
characterized by significant uncertainty on the possible outcomes of
the percolation process. Further, we propose a score, named safeguard
centrality (SC), to identify the nodes that 
have major influence in safeguarding the MCGC at criticality.
We find that the set of top nodes according to SC has a very significant 
overlap with the sets identified as solutions to 
the optimal percolation problem.

\section{Percolation of interdependent multiplex networks}

We consider a multiplex network $\vec{G}=(G^{[1]},G^{[2]})$ formed by $M=2$ layers and $N$ nodes \cite{PhysRep,PRE}.  Each layer $\alpha=1,2$ consists of a network $G^{[\alpha]}=(V,E^{[\alpha]})$. The set $V$ of $N$ nodes  is identical for both layers. The set of links $E^{[\alpha]}$ is instead typical of the layer $\alpha$. 
 We monitor the connectedness of the interdependent multiplex network by looking at the size of the Mutually Connected Giant Component (MCGC)~\cite{Havlin}. The MCGC is the giant component of the multiplex network formed by the largest set of nodes in which each pair of nodes is connected by at least a path in each layer of the multiplex networks (where all these paths must remain inside the MCGC)~\cite{Havlin,Dorogovtsev}.
 
  {In an infinite multiplex network, the MCGC is an extensive
   component in the sense that it includes a non-vanishing fraction of all
   the nodes. The same exact definition doesn't apply to finite real
   networks. However, it is 
usual practice in the field to use the expression MCGC 
in a finite multiplex network to indicate its largest mutually connected 
component. We will interpret this component 
as the {\em giant} one only when its size exceeds $\sqrt{N}$.}

To study the robustness of a multiplex network, we employ a generalized percolation model where 
nodes are initially damaged with probability $f=1-p$ and the relative size $R$ of the MGCC is
monitored as a function of $p$~\cite{Havlin}. The characterization of the robustness 
of the multiplex network thus reduces to the study of the generalized percolation transition.
On infinite networks, the transition is investigated by studying the
average fraction $\bar{R}$ of nodes in the MCGC as a function of
$p$. This critical phenomenon displays noticeable
properties~\cite{Havlin,Dorogovtsev}. The MCGC emerges with a
discontinuous hybrid phase transition at $p=p_c$ where the multiplex
network is affected by avalanches of failures propagating back and
forth among the different layers. This transition has been fully
characterized on multiplex networks with Poisson and scale-free degree
distributions  without edge overlap \cite{Havlin,Dorogovtsev}. In
particular, 
the transition is always discontinuous
and hybrid, and interdependent multiplex networks 
are significantly more fragile than their single layers taken in
isolation \cite{Havlin,Dorogovtsev, Bianconi_book}. 
Recently, it has been shown also that multiplex network models with edge overlap although they tend to be somewhat more robust than multiplex networks without overlap, they present always discontinuous hybrid phase transitions \cite{Goh_comment,Baxter2016,Cellai2016}.

The percolation model applied to multiplex networks  is particularly
relevant in robustness studies of real  interdependent multiplex
networks \cite{Radicchi2015,BianconiRadicchi}. 
However, in a large variety of cases, multiplex networks are far from 
the large network limit. 
It is therefore essential to understand 
whether the average fraction $\bar{R}$ of nodes in the MCGC is 
a suitable metric to assess 
the robustness of real interdependent multiplex networks.

\section{Why a large deviation approach to percolation is needed}
 {
Percolation theory should provide us with
a good prediction of the behavior of the size of the MCGC 
as a function of the probability $f=1-p$ of damaging nodes at random. 
The prediction should include a good estimate of the percolation
 threshold $p_c$. Knowing the value of this quantity is in fact very
 informative as it enables  us to judge  whether the considered
 multiplex network displays a
nonvanishing MCGC or not. Also, the prediction should inform us 
about the nature of the percolation transition, whether this is smooth or
abrupt. The simple analysis conducted below reveals that 
the mean-field theory of
percolation, when applied to real multiplex datasets, 
may fail to provide us with reliable information, as the average
fraction  of  nodes $\bar{R}$ in the MCGC is a metric that does not satisfy the desired
requirements of an informative percolation theory. 
 
\begin{figure*}[!htb]
\centering
\includegraphics[width=0.9\linewidth]{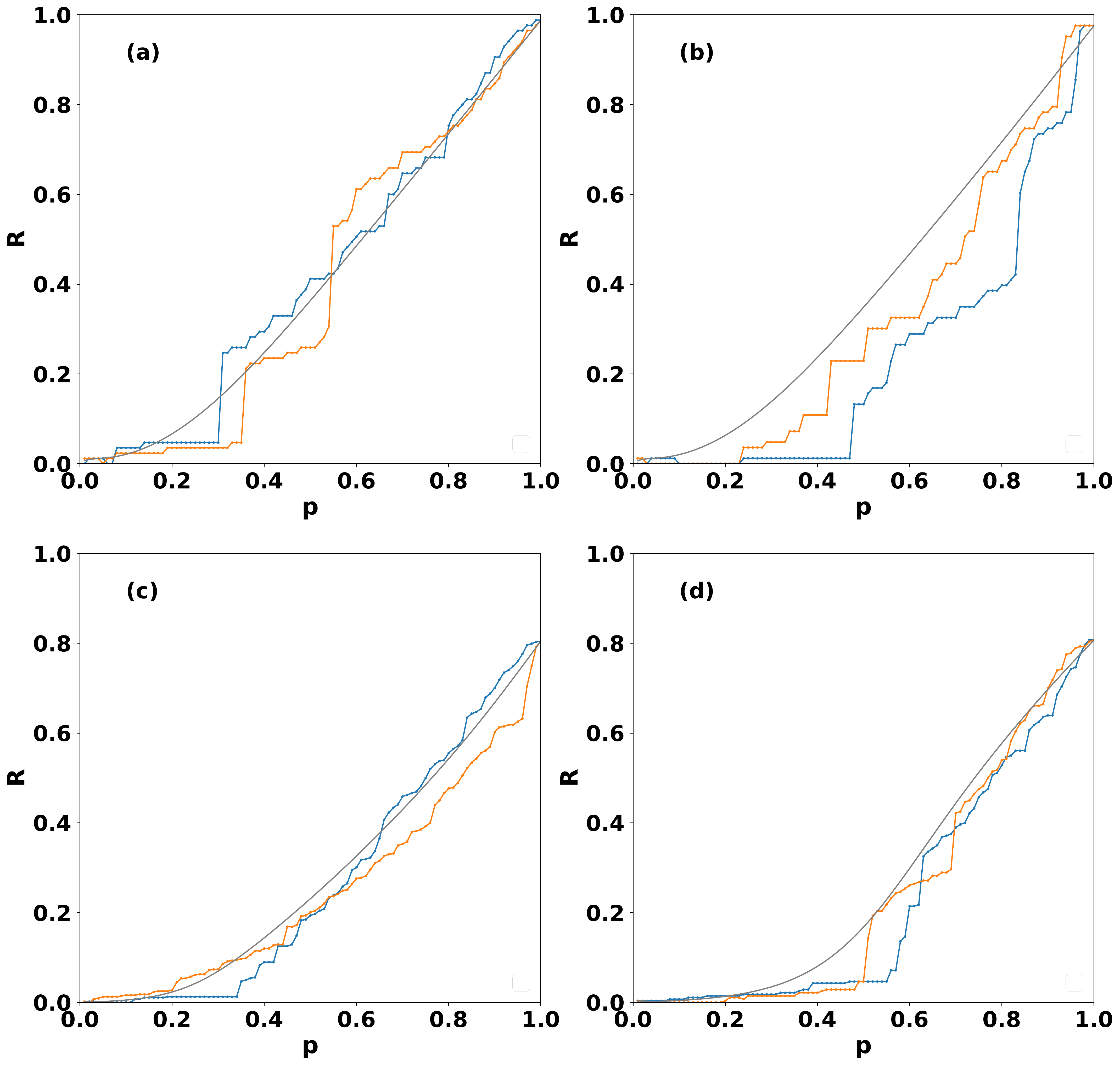}
\caption{{\bf The mean-field approach to percolation can be misleading
    for characterizing the robustness of finite multiplex networks.}
  Single realizations of the percolation process described by the
  corresponding size $R$ of the MCGC as a function of $p$ (blue and
  orange curves) 
are plotted together  with the average size $\bar{R}$ or the MCGC
measured over $10^6$ 
realizations of the initial damage. The different panels correspond 
to four different datasets: the American Airlines-Delta Airlines multiplex network (panel a), Delta
  Airlines-United Airlines multiplex network (panel b),  the Drosophila Melanogaster 
genetic network (panel c), and the C. Elegans connectome (panel d).}
\label{fig0}
\end{figure*}
In analysis, we considered several real-world multiplex networks, including 
air transportation networks among major US air carriers 
(Delta Airlines-United Airlines; American Airlines-United Airlines and United Airlines-Delta
Airlines)~\cite{Radicchi2015}, and biological networks
(the genomic network of the D. Melanogaster \cite{Manlio} and the
C. Elegans connectome \cite{Connectome,Manlio}).
Basic properties of these multiplex networks are reported in Table
$\ref{table1}$.
Number of nodes $N$ range between $73$ and $557$, thus showing that
multiplex networks of practical interest may be
small/medium sized systems. Further,  the comparison between number 
of links $L^{[\alpha]}$ in each layer $\alpha$ and  
total number of multilinks \cite{PRE}
$L^{(1,0)},L^{(0,1)},L^{(1,1)}$,  {
indicating the pair of nodes connected only in layer 1 ($L^{(1,0)}$) only in layer 2 ($L^{(0,1)}$) or in both layers ($L^{(1,1)}$), emphasizes
that the level of link overlap in real multiplex networks may vary
from system to system.}
We simulated a single realization of the random percolation model 
by damaging nodes sequentially with increasing probability $f=1-p$.
In a single realization, we assign a random uniformly distributed variable $x_i$
to each node $i$ of the multiplex networks, and, for each value of $f$, 
we damage all nodes $i$ such that $x_i\leq f $. A single realization of 
percolation can be described by the dependence of  
the relative size $R$ of the MCGC as a function of $p$.

In Figure $\ref{fig0}$, we provide evidence of the large fluctuations
observed for single realizations 
of percolation. The inaccuracy of the 
mean-field approach to percolation in predicting the robustness of
real multiplex  networks of small/medium size is apparent.
For each of the considered datasets, we compare the average 
size $\bar{R}$ of the MCGC and the size $R$ of the MCGC
of two single realizations of percolation. 
As the figure shows, $\bar{R}$ performs poorly with respect to the informative
requirements that we mentioned above.
First, the size $R$ of the MCGC for a given value of $p$ has strong
fluctuations around the mean $\bar{R}$. Second, 
the position of the percolation threshold $p_c$ inferred from 
$\bar{R}$ can be dramatically misleading, providing only a lower bound to the actual transition
points observed in single realizations of the model. 
This fact arises because, for any given $p$, $\bar{R}$ is positive also
when in most of the realizations the size of the MCGC is zero, 
being $\bar{R}$  the average value of non-negative 
numbers. Finally, the abrupt nature of the transitions
associated with individual realizzations is not well captured by 
$\bar{R}$, which instead displays a continuous behavior.

In summary, the mean-field observable $\bar{R}$ underestimates the true fragility
of a real multiplex network.  Correct predictions can be achieved only with an
approach that actually accounts for large deviations. 
}

\begin{table*} \centering
\ra{1.6}
\begin{tabular}{@{}|ccccccc|@{}}\toprule
\hline
\hline
Duplex & N & $L^{[1]}$ & $L^{[2]}$ & $L^{(1,0)}$ & $L^{(0,1)}$ & $L^{(1,1)}$ \\
\hline
American-United airlines & 73 & 229 & 270 & 161 & 202  & 68 \\
American-Delta airlines & 84 & 258 & 442 & 190 & 374  & 68 \\
United-Delta airlines & 82 & 282 & 404 & 226 & 348 & 56 \\
D. Melanogaster genomic network & 557 & 1421 & 1164 & 953 & 696 & 468 \\
C. Elegans connectome & 279 & 514 & 888 & 403 & 777 & 111  \\
\hline
\hline
\bottomrule
\end{tabular}
\caption{{\bf Main properties of the studied datasets.} For each analysed dataset we indicate: the total number of nodes $N$, the total number of links $L^{[1]}$ in layer 1, the total number of links $L^{[2]}$ in layer 2  and the total number of  multilinks $L^{(1,0)},L^{(0,1)},L^{(1,1)}$ indicating the number of pairs of nodes connected only in layer 1, only in layer 2 or in both layers, respectively.}
\label{table1}
\end{table*}
\section{Large deviation approach to percolation}

 In this paragraph, we establish  the general theoretical framework for 
characterizing the large deviation properties of percolation in interdependent 
multiplex networks. Our goal is to  
quantify the response of a multiplex network to an initial damage of 
the nodes using a metric different from the  
mere average fraction $\bar{R}$ of 
nodes in the MCGC. In particular, we 
will explore the properties  of  the entire  distribution $\pi(R)$ of 
observing a MCGC formed by a fraction $R$ of nodes.
The distribution $\pi(R)$ will be studied as a function of $p$, i.e., 
the probability that a node is not initially damaged.

\begin{figure*}[!htb]
\centering
\includegraphics[width=0.9\linewidth]{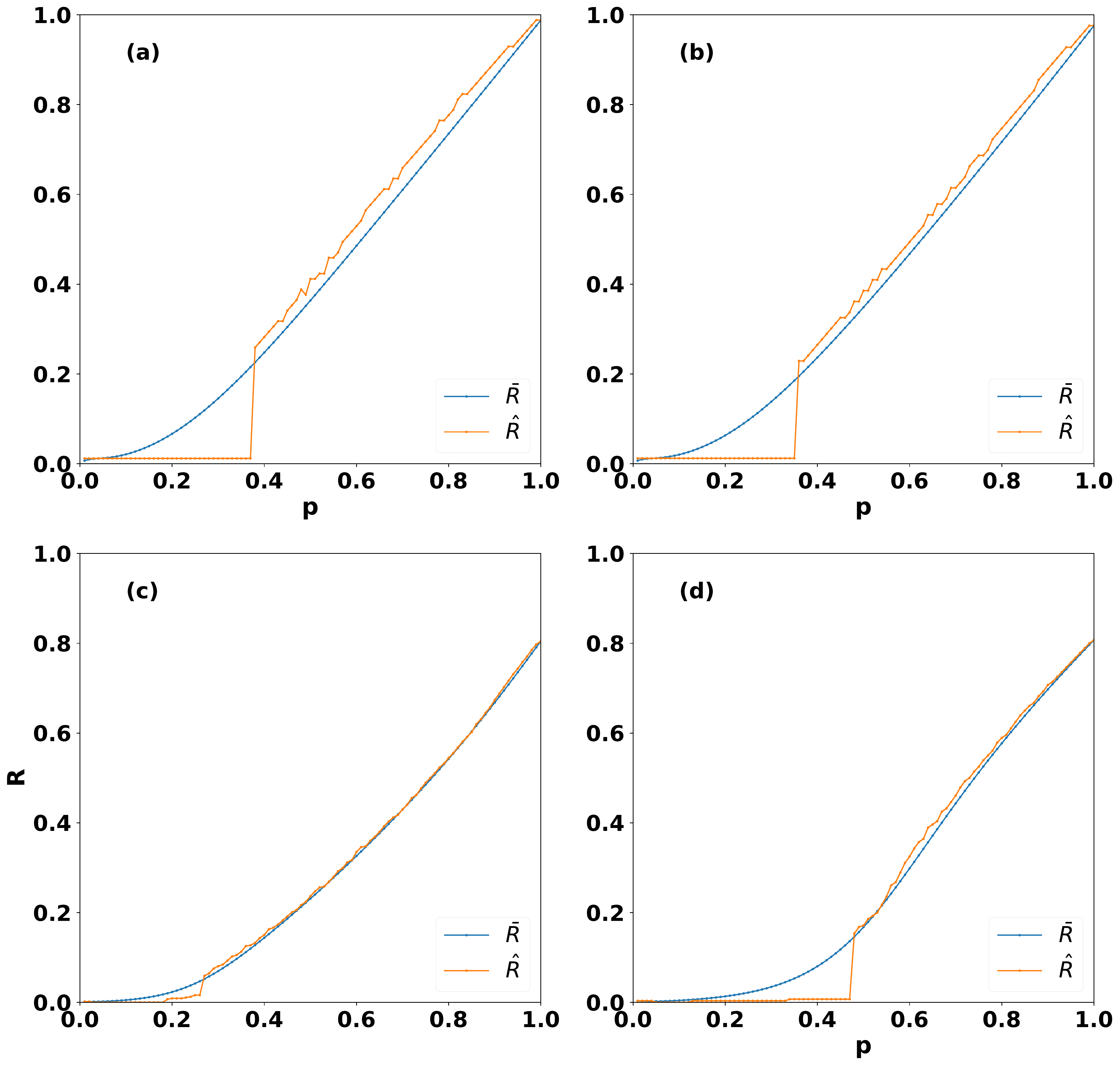}
\label{}
\caption{{\bf Typical versus average size of the MCGC.} We report
  $\bar{R}$ and $\hat{R}$ for four different data sets: the American Airlines-Delta Airlines multiplex network (panel a) Delta
  Airlines-United Airlines multiplex network (panel b),  the Drosophila Melanogaster 
genetic network (panel c)  the C. Elegans connectome (panel d). The
  curves $\bar{R}$ vs. $p$ and $\hat{R}$ vs. $p$ have been numerically
  calculated from the distributions $\pi(R)$ of observing a MCGC with
  relative size $R$.
The distribution $\pi(R)$ is  constructed for each value of $p$ by
performing $P$ initial realizations of the damage.
We use $Q=10^6$ for all the multiplex  network datasets. 
 }
 \label{figure1}
\end{figure*}
We consider a large number $Q$ of random initial damage realizations.
Each initial damage configuration $\mu=1,2,\ldots, Q$ is denoted 
by  $\{s_i^{\mu}\}_{i=1,2,\ldots, N}$, where 
$s_i^{\mu}=0$ if node $i$ is initially damaged, and
$s_i^{\mu}=1$, otherwise.  
We assume that each node is damaged independently
with probability $f=1-p$. 
Therefore, the probability associated to the initial damage realization $\{s_i^{\mu}\}$ is 
\bea
{\mathcal P}(\{s_i^{\mu}\})=\prod_{i=1}^N \left[ps_i^{\mu}+(1-p)(1-s_i^{\mu})\right].
\eea

For each initial damage configuration,  
we determine whether node $i$
belongs to 
the MCGC, i.e., $\sigma_i^{\mu} = 1$,  or not, i.e.,  
$\sigma_i^{\mu} = 0$. The fraction  $R^{\mu}$  
of nodes in the MCGC is 
\bea
R^{\mu}=\frac{1}{N} \sum_{i=1}^N \sigma_i^{\mu}.
\eea
For any given value of $p$, different initial damage configurations
might induce  
MCGCs of different sizes.
In order to study the distribution $\pi(R)$ of the fraction of the
nodes in the 
MCGC for a random realization of the initial damage $\{s_i^{\mu}\}$
with 
probability ${\mathcal P}(\{s_i^{\mu}\})$ we consider a large number
$P$ 
of realizations of the initial damage and we estimate $\pi(R)$ as

\bea
\pi(R)=\frac{1}{Q \, (\Delta R)}\sum_{\mu=1}^Q \delta(R,R^{\mu}), 
\eea
where $\delta(x,y) = 1$ if $x=y$, and $\delta(x,y)=0$, otherwise.
 { $\Delta R=1/N$ represents the width of the bins used for
  estimating the distribution, thus providing the proper normalization condition for $\pi(R)$.}
From the full distribution $\pi(R)$ of the sizes $R$ of the MCGCs, 
it is possible to extract two major statistical
quantities: 
the average size of the MCGC, namely $\bar{R}$, and the typical 
(most probable) size of the MCGC, namely $\hat{R}$. The quantities are
defined respectively as 
\bea
\bar{R}&=&\frac{1}{N} \, \sum_{R} R\  \pi(R) 
\eea
and
\bea
\hat{R}&=& \arg \; \max_{R}\; [ \pi(R) ] .
\eea

We stress once more that all quantities defined above are defined
given the probability $f=1-p$  for the random initial damage of each
node. We avoid to write explicitly such a 
dependence just for  shortness of notation.

\section{Large deviation of percolation in real multiplex networks}
\subsection{Typical versus average size of the MCGC}

In order to explore the large deviation properties of percolation on real multiplex network
we considered again the real-world multiplex networks listed in Table
$\ref{table1}$.
We have calculated $\pi(R), \bar{R}, \hat{R} $ by performing
numerically 
$Q=10^6$ realizations of the initial damage as a function of $p$.
Our results  reveal that  the typical response to damage $\hat{R}$
uncovers a completely different scenario with respect to the one
indicated by the average response to damage $\bar{R}$ for each of the
studied datasets 
(see Figure $\ref{figure1}$).
Indeed, while $\bar{R}$ decreases smoothly for decreasing values of
$p$, suggesting that the system might be robust to damage, 
$\hat{R}$ reveals a discontinuous behaviour  
with a rapid jump of $\hat{R}$ from $R=R_c\gg 1/N$  to $R=1/N$  {at the {\it effective critical threshold }
$p=p_c$ where $\pi(1/N)=\pi(R_c)$.  }
 {Hence, $\hat{R}$ highlights a  risk of systemic
  failure that is not visible from $\bar{R}$, pointing out  a
  serious shortcoming of the average 
metric in characterizing the true fragility of the system.
}

\begin{figure*}[!htb]
\centering
\includegraphics[width=0.9\linewidth]{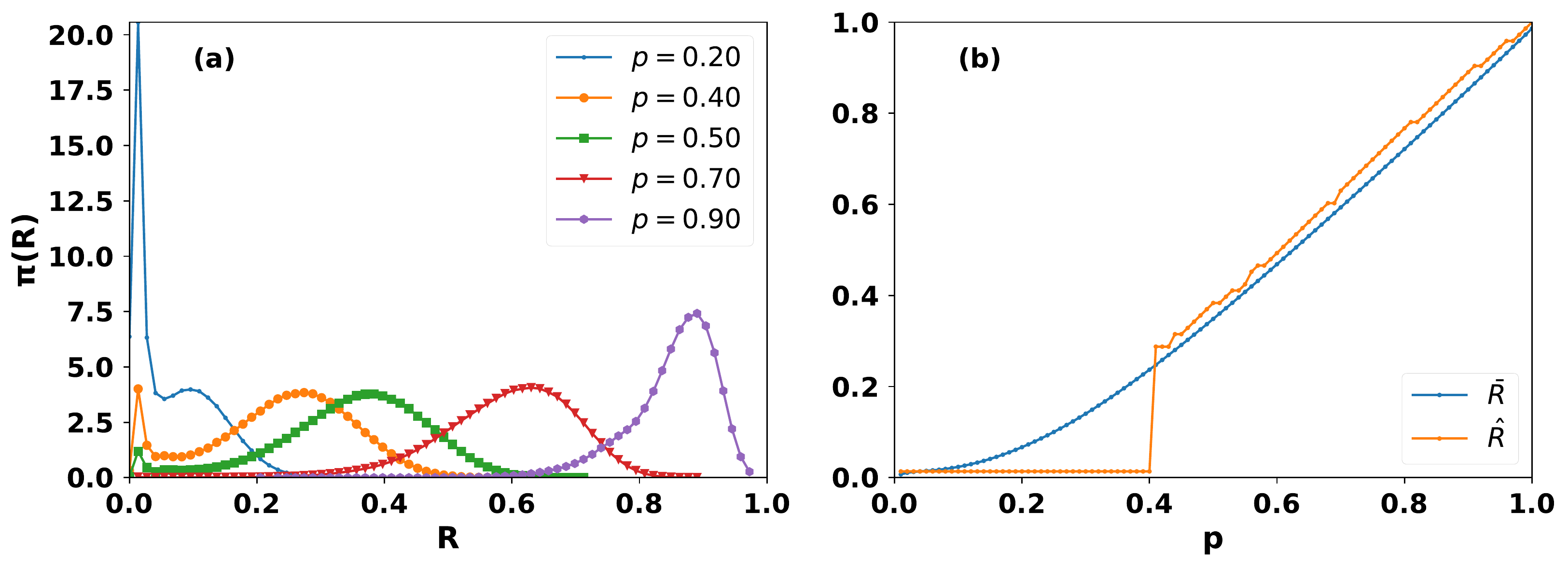}
\caption{{\bf The large deviation study  of percolation on the
    American Airlines-United Airlines duplex network.} The probability
  distribution $\pi(R)$ of observing a MCGC of relative size $R$ in the American Airlines-United Airlines duplex network is shown in panel (a) for different values of $p$. The average $\bar{R}$ and the most likely $\hat{R}$ size of the MCGC of the same dataset are plotted as a function of  $p$ in panel (b). These results are obtained from numerical simulations of $Q=10^6$ random initial realizations of the damage.}
\label{figure2}
\end{figure*}

\subsection{Bimodality of the distribution of the size of the MCGC}

\begin{figure*}[!htb]
\centering
\includegraphics[width=0.9\linewidth]{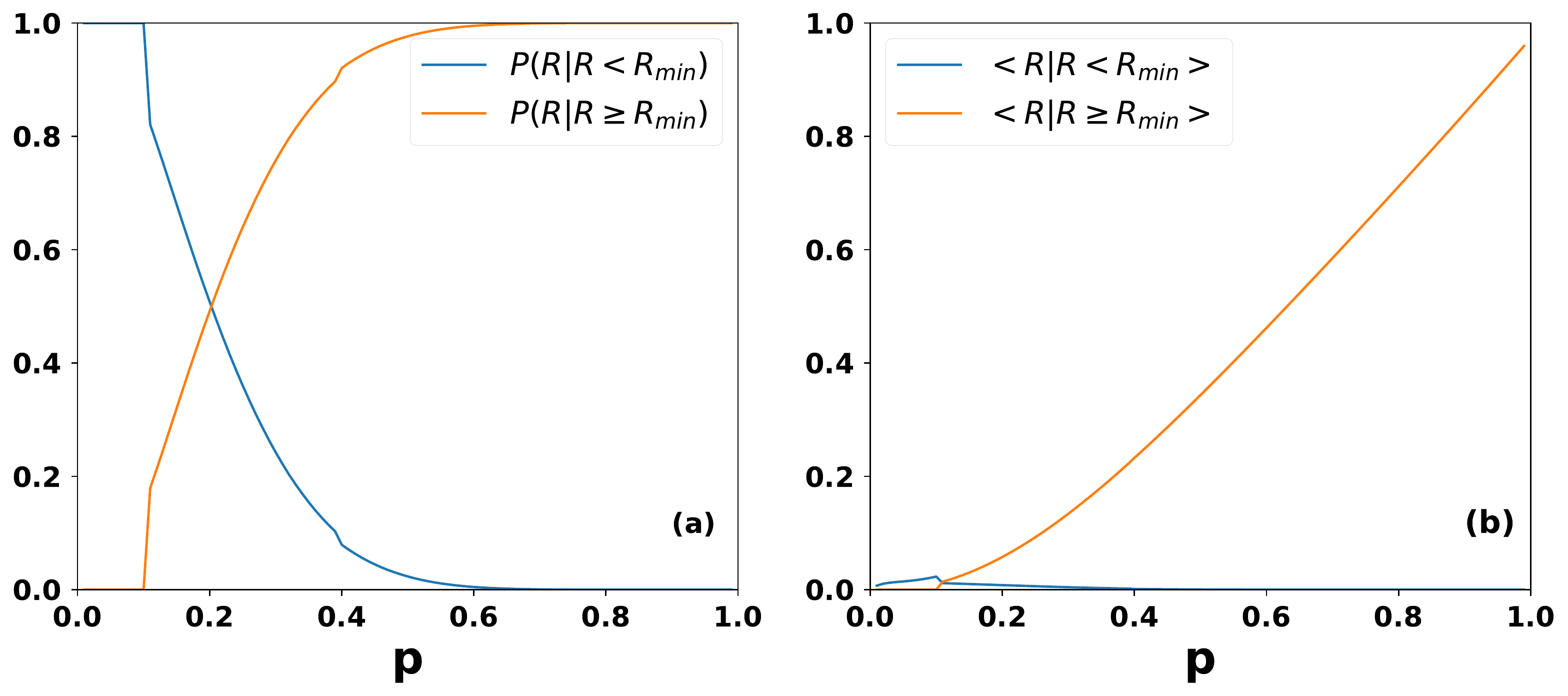}
\caption{{\bf The characterization of the two modes of the distribution $\pi(R)$.} The probabilities $P(R\geq R_{min})$, $P(R< R_{min})$ (panel a) and the average size $\avg{R|R\geq R_{min}}$ and $\avg{R|R>R_{min}}$ of the MCGC corresponding to the two modes (panel b) are plotted as a function of $p$ for the American Airlines-United Airlines duplex network. These results are obtained by performing $Q=10^6$ realizations of the initial damage.}
\label{figure_Smin}
\end{figure*}

 {In order to fully characterize the large deviation properties of
  percolation, we need to study the 
 probability distribution $\pi(R)$ of the size $R$ of the MCGC. 
The characterization of this distribution 
as a function of the probability $p$  
will reveal why $\hat{R}$ has a discontinuous 
behavior as a function of $p$. }
Here we have
considered in detail the 
American Airlines-United Airlines dataset.
For this dataset, the
probability distribution 
$\pi(R)$ can be studied together with $\bar{R}$ and $\hat{R}$ as a
function of $p$ (see Figure $\ref{figure2}$). 
Starting from high values of $p$ and decreasing $p$, we observe that 
initially the distribution $\pi(R)$ is unimodal, and the most likely
outcome $\hat{R}$ decreases. However, for lower values of $p$,  the
distribution $\pi(R)$ becomes bimodal 
and for $p =p_c\simeq 0.40$
 it has two maxima at $R=R_c \simeq 0.27$
and $R=1/N\simeq 0.014$ with $\pi(1/N)=\pi(R_c)$.  
Finally, for even lower values of $p$, i.e., for $p<p_c$,  $R=1/N$
becomes the most likely size of 
the MCGC.
 {We checked that the emergence of a bimodal
  distribution is not an artifact of specific correlations
  build in the multiplex network structure. In fact, we tested
  that the feature of the distribution is unaffected by different randomization
  procedures of the structure of the multiplex (see Supplementary Information \cite{SI}).

We stress that the large deviation of percolation consists in the full 
characterization of the distribution $\pi(R)$, 
and not exclusively on the characterization of the typical size $\hat{R}$ of the MCGC.
If we just based our prediction about the robustness of a real multiplex
network on the observation of the typical size  $\hat{R}$ of the MCGC, we would
overestimate its fragility. In fact, even
for $p < p_c$ so that $\hat{R} = 0$, 
there is a significant probability that the MCGC is
non-vanishing. In this regime, $\pi(R)$ is still 
bimodal, and the width of the distribution around the two peaks is not symmetric.}
 As such, even
if the left peak is higher than the right one, still the
right mode of $\pi(R)$ may have higher weight. 
 {
The relative weight of these two modes is 
clearly an important  information for assessing 
the robustness of the multiplex network as a whole. In  Figure
$\ref{figure_Smin}$, we propose an analysis that accounts for them.
Indicate with $R_{min}$ the position of the local minimum in the
distribution $\pi(R)$ separating the two peaks. Then evaluate the
relative weight of the two modes by measuring the probability
$P(R < R_{min})$ and 
its complementary probability $P(R \geq R_{min})$. 
The metrics reveal that at $p=p_c$ the probability $P(R\geq R_{min})$
is still larger than 50\%, thus providing information that is 
not directly retrievable either from $\hat{R}$ or $\bar{R}$.
In principle, any scalar observable extracted from the distribution $\pi(R)$
can provide new information about the system robustness.
However, we argue that the combined use of $\hat{R}$
and $\bar{R}$ generates already sufficient knowledge for many
practical tasks. These metrics have in fact the nice feature of representing respectively
an over-estimation and under-estimation of the fragility of
the  multiplex network, thus identifying the range of $p$ values 
where  the  percolation transition is expected to happen. 
This interpretation of $\hat{R}$ and $\bar{R}$
will become more precise in Sec VI where  we address finite-size
effects, and we will show 
that in the limit of
infinite networks the two metrics become identical.
}

\begin{figure}[!htb]
\centering
\includegraphics[width=0.8\columnwidth]{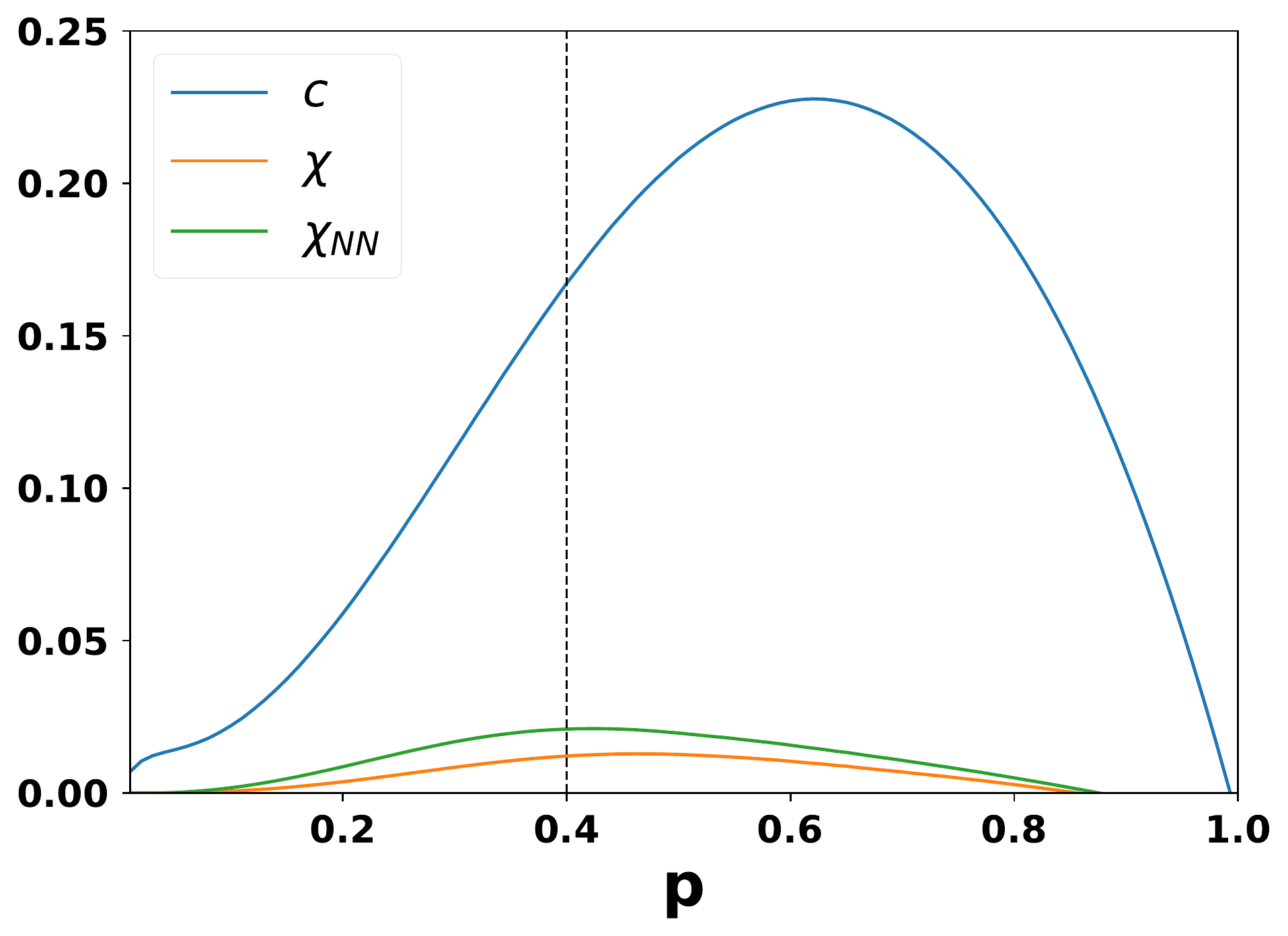}
\caption{
{\bf Correlations and specific heat for the American Airlines-United
  Airlines duplex network.} The specific heat of percolation $C$,  the
correlation coefficient $\chi_{NN}$ among  neighbor nodes,  and  the
correlation coefficient $\chi$ among any pair of nodes calculated for the
American Airlines-United Airlines dataset are plotted as a function of
$p$. The solid dashed line indicates the effective critical point $p=p_c$. 
These results are obtained from numerical simulations of 
$Q=10^6$ random realizations of the initial damage.}
\label{figure3}
\end{figure}

\subsection{Susceptibility, Specific heat and Correlations}

To further characterize the properties of the system, 
we studied
correlations existing among the states of different nodes.
 {This study is interesting from two different points of
  view. First, correlations are not captured by the mean-field
  approach to percolation; hence, their analysis allows us to obtain
information that is usually neglected in standard analyses. Second, 
average correlations between nodes allows us to define a well-grounded 
version of the susceptibility
for percolation on multiplex networks. This definition is in line with
the usual practice adopted in the study of critical phenomena, where
the susceptibility of a system is taken proportional to the average correlation. 

We define  the average correlation $\chi$ between the state 
of every pair of nodes in different realizations of the damage 
and the average correlations   $\chi_{NN}$ calculated only among 
of neighboring nodes in at least one layer of the multiplex
network. These quantities are  defined respectively as}
\bea
\chi&=&\frac{1}{N(N-1)}\sum_{i\neq j}\left[\Avg{\sigma_i\sigma_j}-\avg{\sigma_i}\avg{\sigma_j}\right],\nonumber\\
\chi_{NN}&=&\frac{1}{\avg{k}N}\sum_{i=1}^N\left[\sum_{j \in \mathcal{N}_i}\left[\Avg{\sigma_i\sigma_j}-\avg{\sigma_i}\avg{\sigma_j}\right]\right],
\eea
where we indicated with $\Avg{\sigma_i}$ and $\Avg{\sigma_i\sigma_j}$ the averages
\bea
\Avg{\sigma_i}&=&\frac{1}{Q}\sum_{\mu=1}^Q \sigma_i^{\mu},\nonumber \\
\Avg{\sigma_i\sigma_j}&=&\frac{1}{Q}\sum_{\mu=1}^Q
\sigma_i^{\mu}\sigma_j^{\mu} , 
\eea
and we indicated with $\mathcal{N}_i$ the set composed by all neighbors
of node $i$ in at least one layer of the multiplex network. 
 Moreover, we have evaluated the recently introduced {\em specific heat} $C$  of percolation \cite{Fluct}
with $C=Nc$ and $c$ defined as 
  \bea
	c&=&\frac{1}{N}\sum_{i=1}^N\avg{\sigma_i}(1-\Avg{\sigma_i})
  \eea
  indicating the average fluctuations of the state of a single node.
The specific heat $C$ together with the correlation coefficient $\chi$ determines the variance $\sigma^2_R$ of the size of the giant component $R$. In fact we have   
\bea
\sigma^2_R&=&\frac{1}{N^2}\sum_{i,j}\left[\Avg{\sigma_i \sigma_j}-\sum_{i,j}\avg{\sigma_i}\avg{\sigma_j}\right] \nonumber \\
&=& \chi \left(1-\frac{1}{N}\right) + \frac{C}{N^2}.
\eea
In Figure $\ref{figure3}$,  we plot $c, \chi$ and $\chi_{NN}$ as a function of $p$ for the American Airlines-United Airlines duplex network.
From this figure, it is possible to show that these curves display a
maximum as a function of $p$. We note that for very small values of
$p$, when the MCGC is also very small, the correlation coefficients and the
specific heat are expected to be small since typically most of the
nodes will be damaged. Similarly when $p$ is approaching one, most of
the nodes will be undamaged yielding small correlations and specific
heat. Therefore, the observed  maximum of $c, \chi$ and $\chi_{NN}$ as
a function of $p$ is  expected. 

 { For a good susceptibility measure, 
we would like to observe a maximum in correspondence of the
percolation threshold.  We
note that the maximum of  $\chi_{NN}$ is achieved for 
values of $p$  that are closer to the transition point $p=p_c$ than
those corresponding to
the maximum of $\chi$. Additionally from Figure $\ref{figure3}$,
we can notice that correlations among nearest neighbors are, on
average, higher than the correlations among 
any pair of nodes, i.e. $\chi_{NN}\geq\chi$.
 }

\subsection*{Overlap between MCGCs}
 {
 We have emphasized that, on finite networks, 
MCGCs resulting from two initial damage configurations    
drawn from the same distribution ${\mathcal P}(\{s_i^{\mu}\})$ can have different relative size $R$.
  In order to  quantitatively evaluate how  similar are two different MCGCs resulting from two different configurations $\mu$ and $\nu$ of the initial damage   we propose to use the {\it overlap} $q^{\mu,\nu}$.
 The overlap $q^{\mu,\nu}$ is given by the sum between  fraction of nodes that belong to both MCGCs and the sum of nodes that do not belong to the MCGC for both realizations $\mu$ and $\nu$ of the initial damage, i.e.,
\bea
q^{\mu,\nu} = \frac{1}{N} \sum_{i=1}^N \left[\sigma_i^{\mu}
  \sigma_i^{\nu} + (1-\sigma_i^{\mu}) (1-\sigma_i^{\nu})\right] \; , 
\eea
where $\sigma_i^{\tilde{\mu}}=1$ ($\sigma_i^{\tilde{\mu}}=0$) indicates that node $i$ is in (is not in) the MCGC after the initial damage configuration $\tilde{\mu}$ with $\tilde{\mu}\in\{\mu,\nu\}$.
By definition, we have that
$q^{\mu,\nu}\in [0,1]$. Values of overlap close to one indicate that  
the two MCGCs have a very similar node composition, while values of
the overlap close to zero indicate that the two MCGCs are very
different in terms of node composition. For values of $p$ close to one,
where most of the nodes belong to the MCGC, and for values of $p$ close
to zero, where most of the nodes do not belong to the MCGC, the typical
overlap among MCGCs is expected to  be high; instead, typical overlap
values should be small for intermediate values of $p$. 
We evaluated 
the  average value $\bar{q}$  and the standard deviation
$\sigma_{\bar{q}}$ 
of the overlap measured 
for different values of $p$.
These metrics are computed using $\tilde{Q}$ pairs of realizations of the 
initial damage  $(\mu_n,\nu_n)$ (with $0<n\leq \tilde{Q}$), and using the 
definitions
\bea
\bar{q}&=&\frac{1}{\tilde{Q}}\sum_{n=1}^{\tilde{Q}}q^{\mu_n,\nu_n},\nonumber \\
\sigma_{\bar{q}}&=&\frac{1}{\tilde{Q}}\sum_{n=1}^{\tilde{Q}}\left(q^{\mu_n,\nu_n}-\bar{q}\right)^2.
\eea

\begin{figure}[!htb]
\centering
\includegraphics[width=0.9\linewidth]{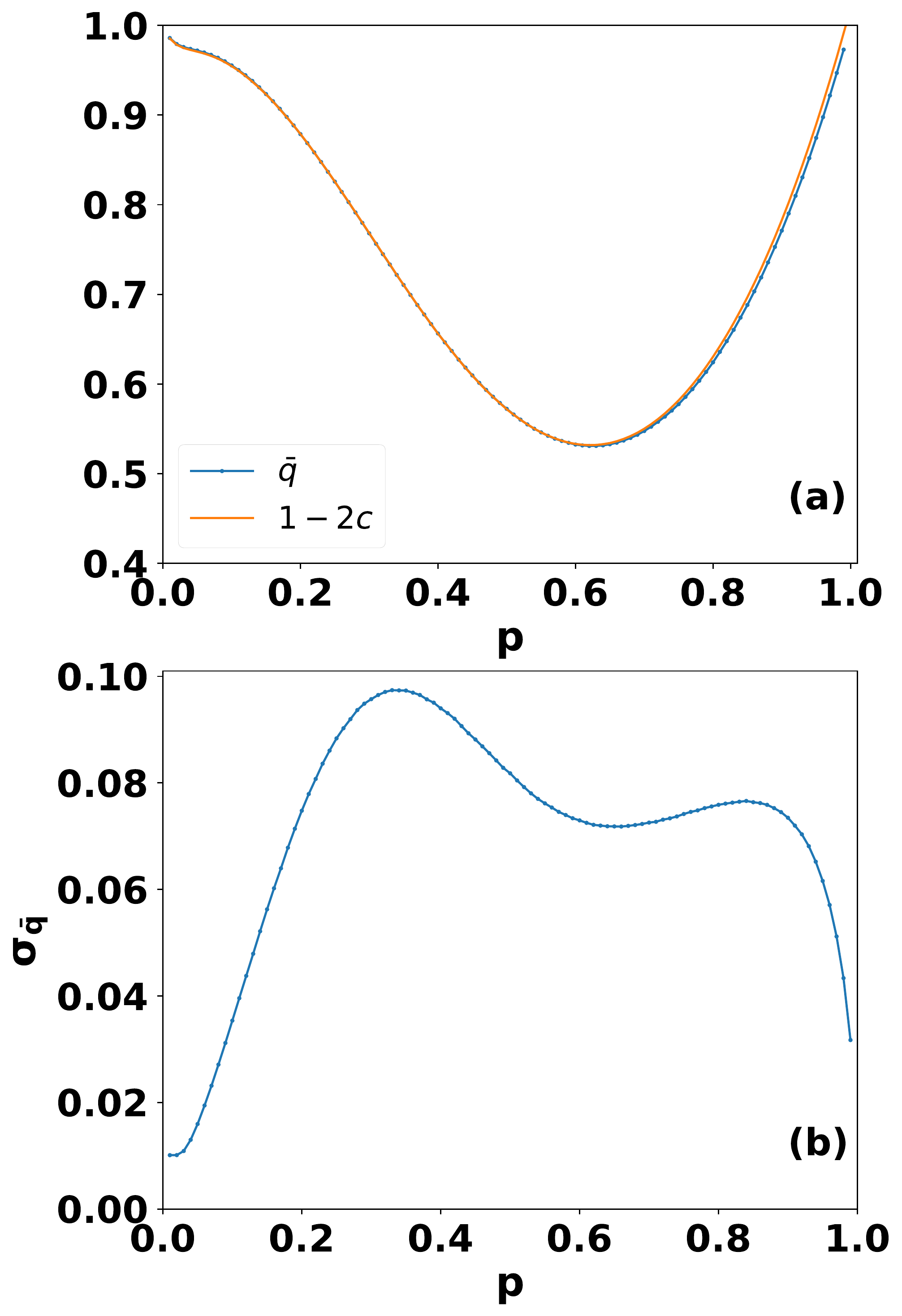}
\caption{The mean value $\bar{q}$  (panel a) and the  standard deviation $\sigma_{\bar{q}}$ (panel b) of the overlap distribution $\rho(q)$ are plotted versus $p$ for the American Airlines-United Airlines multiplex network. 
These statistical properties have been numerically evaluated starting from $\tilde{Q}=10^6$ pairs of random realizations of the initial damage. }
\label{figureS4}
\end{figure}
\begin{figure}[!htb]
\centering
\includegraphics[width=0.9\linewidth]{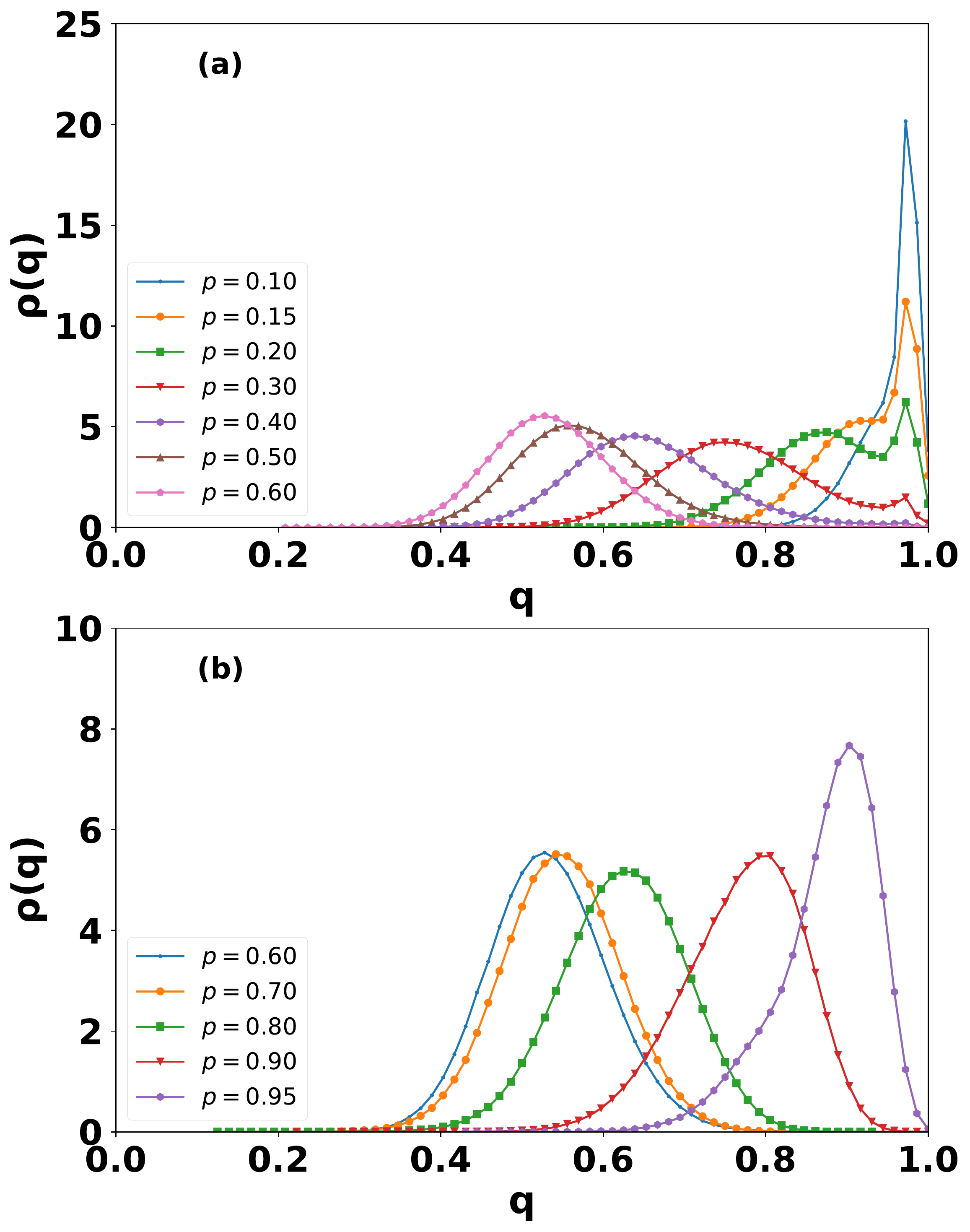}
\caption{The overlap distributions $\rho(q)$  among the MCGCs of the American Airlines-United Airlines duplex network are plotted for values of $p$ given by $p=0.1,0.2,0.3,0.4,0.5,0.6$ (panel a) and $p=0.6,0.7,0.7,0.9,0.99$ (panel b).  These results indicate that for $p<0.6$   the typical overlap among MCGCs decreases with increasing values of $p$ while for $p>0.6$ the typical overlap among MCGCs increases as $p$ increases. This distributions are calculated starting from $\tilde{Q}=10^6$ pairs of random realizations of the initial damage. }
\label{figureS3}
\end{figure}
We note that $\bar{q}$ is expected to be strongly correlated with the specific heat $C=Nc$~\cite{Fluct}.
In fact,  for large values of $\tilde{Q}$, 
we can approximate $\bar{q}$ with 
\bea
\bar{q}&\simeq &\frac{1}{N}\sum_{i=1}^N \left[\Avg{\sigma_i}^2+(1-\Avg{\sigma_i})^2\right]\nonumber \\&=&1-2c.
\label{qc}
\eea
In Figure $\ref{figureS4}$, we report $\bar{q}$ and $\sigma_{\bar{q}}$ calculated over $\tilde{Q}=10^6$ pairs of random realizations of the initial damage performed over the American Airlines-United Airlines multiplex networks. We observe that Eq. (\ref{qc}) is satisfied. Therefore, $\bar{q}$ has a minimum corresponding to the maximum of $c$. Interestingly, $\sigma_{\bar{q}}$ displays two maxima as a function of $p$, and achieves its absolute maximum for values of $p$ preceding $p=p_c=0.4$, i.e., $p=0.34$.  

Nevertheless, the full distribution of the overlap $\rho(q)$ observed
over $\tilde{Q}$ pairs of realizations of 
the initial damage encodes more information than its average
$\bar{q}$. In particular, the distribution 
reflects the many-body correlations existing among the state of different nodes (see Figure $\ref{figureS3}$). 
We note that for $p<p_c$, in correspondence of the bimodal regime for
the distribution $\pi(R)$, also $\rho(q)$ is
bimodal, with a second peak at high values of the overlap that reflect 
configurations where the network is completely dismantled.
Deviations of the distribution $\rho(q)$ from a single peak clearly
indicate the complex many-body 
interactions present in the system, 
and reveals the important effect of the fluctuations of the MCGC 
in a way that is reminiscent of phenomena observed in disordered systems \cite{Parisi_Mezard}.
}
\section{Finite-size effects}
The discrepancy between $\hat{R}$ and $\bar{R}$ is an effect of the finite size of the networks
analyzed. For an infinite network, the percolation transition is known to be
self-averaging, i.e., the difference between $\bar{R}$ 
and $\hat{R}$ is vanishing.
To explore for which network sizes we should expect significant
differences between $\hat{R}$ and $\hat{R}$, we performed a large
deviation study of percolation on synthetic multiplex networks.
We considered duplex 
networks of sizes $N=10^2,10^3,10^4$ in 
which each layer is a random network with  
Poisson degree distribution and average degree $z=5$.
We observe that, as $N$ increases, the percolation
transition becomes self-averaging, and that $\bar{R}$ approximates
increasingly 
better $\hat{R}$ (see Figure $\ref{figure4}$)  {, and the distribution $\pi(R)$ becomes increasingly narrow (see Supplementary Information \cite{SI}).
Therefore, the distribution $\pi(R)$ is bimodal  for a range of values
of $p$ that  is converging to zero  as $N$ increases. 
Characterizing the large deviation of percolation is particularly important for investigating the robustness of small/medium size multiplex networks as the ones considered  in this study.}
\begin{figure}[!htb]
\centering
\includegraphics[width=0.9\columnwidth]{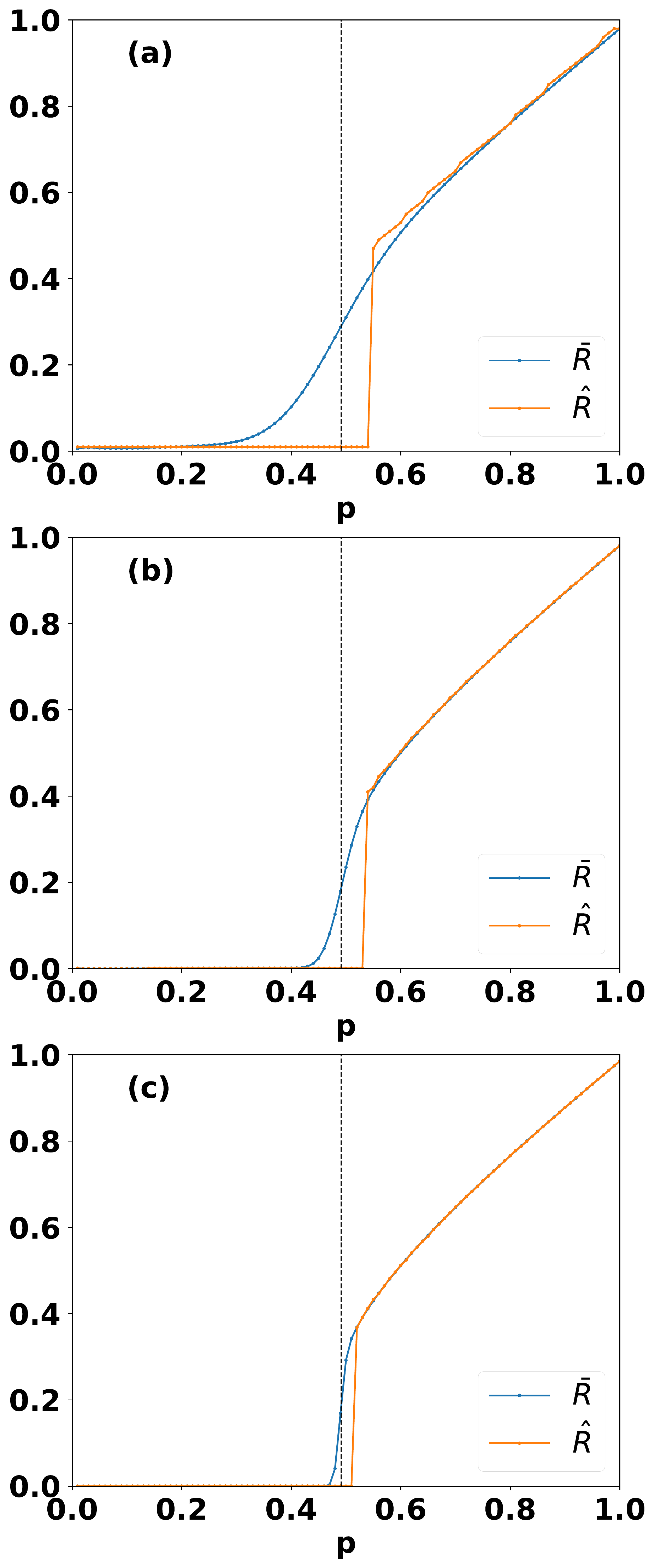}
\caption{{\bf Finite-size effects of percolation on a Poisson duplex network.} The distribution $\pi(R)$ of the size of the MCGC, the average  $\bar{R}$ and the typical  $\hat{R}$ size of the MCGC are plotted versus $p$ for  Poisson duplex networks with average degree $z=5$ and network sizes $N=10^2$ (panel a), $N=10^3$ (panel b), $N=10^4$ (panel c).  This numerical results are obtained  by running  $Q=10^6$ (panel a,b) and $Q=10^5$ (panel c) realizations of the initial damage configurations.  The dashed solid line indicates the theoretically predicted percolation threshold in the limit $N\to \infty$. }
\label{figure4}
\end{figure}
Interestingly,  the average response to damage $\bar{R}$ and the
typical response to damage $\hat{R}$ differ significantly up to
network sizes of several thousand of nodes. Duplex networks of these
size are very common,  and include not only brain networks and
air transportation networks such as those studied here, but also
interdependent power-grids, ecological multiplex networks and brain
functional networks. We believe therefore that our results might be 
relevant for scientists investigating the robustness of very 
different types of real multiplex datasets.

\section{Safeguarding the MCGC}

Above, we defined the critical point $p_c$ as the value of $p$
where the two peaks of the bimodal distribution $\pi(R)$ have the same
height. For $p=p_c$, we have that the left peak is located at $R=1/N$, while the
right peak is located at $R=R_c$, with $R_c \gg 1/N$.
The condition $\pi(R=1/N) = \pi(R=R_c)$ tells us that the likelihood that
the system fails is  comparable with the probability  that the
system is still in the functioning state. Is it possible to predict
initial configurations of damage that lead to one or the other
final states of the system?  Is it possible to safeguard some nodes so
that the sufficient condition that the network will be in the
functional state is met? Please note that the latter question is 
different from the one defined in optimal percolation, where the goal
is to dismantle a system rather than preserving its
cohesiveness~\cite{Makse, Radicchi2017}.

Here, we propose an  
algorithm that ranks nodes according to their influence
 in determining the size of the MCGC.  The algorithm uses the
 bimodality of $\pi(R)$, and is designed to be effective for $p=p_c$.
We name the score resulting from the algorithm as
{\it safeguard centrality}. 
The algorithm starts by defining  two ranges 
of possible sizes for the  MCGCs,  {either a well-defined MCGC
  $R>R^{\star}=\frac{1}{\sqrt{N}}$ 
or a dismantled network with $R<R^{\star}$.} A score $\Delta s_i$ is assigned to every node $i$.
$\Delta s_i$ is defined as the difference
between the joint probability  that node $i$ is not initially 
damaged and
$R>R^{\star}$, and the joint probability that node is not   
initially damaged and $R<R^{\star}$, i.e., 
\bea
\Delta s_i=\frac{1}{P}\sum_{\mu=1}^P s_i^{\mu}[\theta(R^\mu-R^{\star})-\theta(R^{\star}-R^\mu)],
\eea
where  $\theta(x)$ is the Heaviside function, i.e.,
$\theta(x) = 0$ if $x<0$,  and
$\theta(x) = 1$, otherwise.
Nodes with top $\Delta s$ values are nodes whose safeguard
may result in large MCGC sizes, i.e., the nodes that are responsible 
for keeping the multiplex connected.

\begin{figure*}[!htb]
\centering
\includegraphics[width=0.7\linewidth]{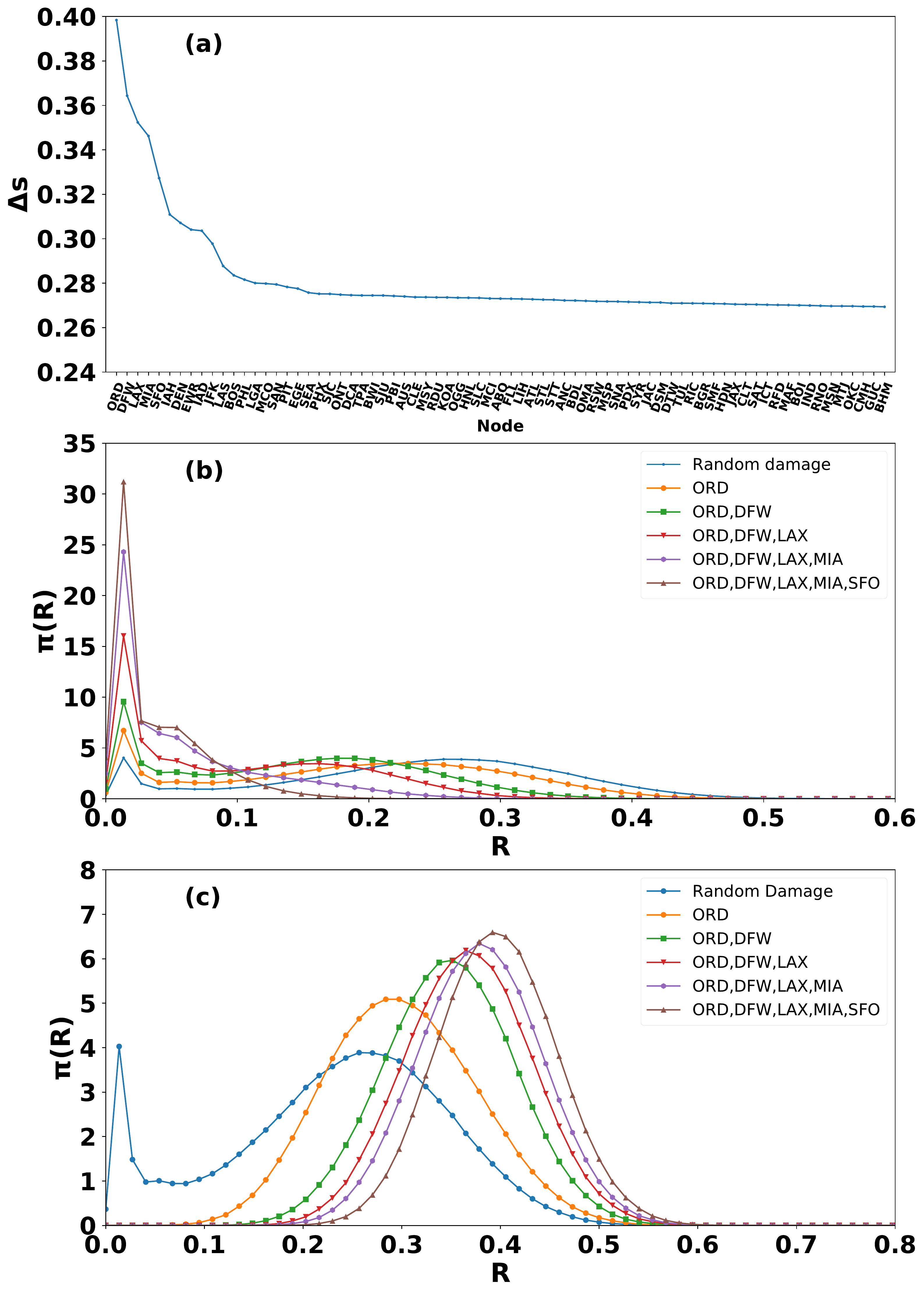}
\caption{{\bf Effect of the damage and the safeguard of the top-ranked nodes on the robustness of the American Airlines-United Airlines duplex network.} The centrality measure $\Delta s$ for each of the $N=73$ airports of the American Airlines-United Airlines dataset is shown in panel (a). The centrality measures is evaluated by considering $P=10^6$ realization of the initial damage at $p=p_c=0.40$ taking  $R^{\star}=1/\sqrt{N}<R_c=0.27$.   
The distribution $\pi(R)$ of the size $R$  of the MCGC at $p=p_c$ is compared to the distribution $\pi(R)$ obtained when the top-ranked nodes according to $\Delta s$ are damaged for sure (panel (b)) or safeguarded (panel (c)) while the other nodes are damaged with probability $p=p_c=0.40$. The distribution in panels (b) and (c) are obtained considering $10^6$ realizations of the initial damage.}
\label{figure5}
\end{figure*}

\begin{table}[!htb] \centering
\ra{1.6}
\begin{tabular}{@{}|ccccccccc|@{}}\toprule
\hline
\hline
Airport & $\Delta s$ & SA & HDp & HDs & HDAp & HDAs & CI1p & CI1s\\
\hline
ORD & 0.3984  & 0 & 1 & 1 & 1 & 1 & 1 & 1\\
DFW & 0.3644 & 1 & 3 & 2 & 3 & 2 & 8 & 6\\
LAX & 0.3523 & 0 & 2 & 5 & 2 & 5 & 2 & 5\\
MIA & 0.3462 & 1 & 7 & 7 & 8 & 7 & 9 & 9\\
SFO & 0.3273 & 0 & 4 & 6 & 5 & 6 & 3 & 4\\
\hline
Kendall-$\tau$& 1 & 	0.8  &  0.6   &  0.8    &  0.6     &0.8       &0.4  &0.2 \\
\hline
\hline
\bottomrule
\end{tabular}
\caption{
 {The top 5  airports in the American Airlines-United
  Airlines duplex network according to the centrality measure $\Delta
  s$ are listed together with  their 
corresponding classification $\{s_i^{\star}\}$ 
according to the SA algorithm ($s_i^{\star}=0$, if node $i$ belongs to
the set of structural nodes, $s_i^{\star}=1$, otherwise) and their rank according to the HDp,
HDs, HDAp, HDAs, CI1p and  CI1s algorithms. }The last row indicates the
Kendall-$\tau$ correlations among the ranking of these 5 airports
according to $\Delta s$ and each of the other state of the art
algorithms. Note that when comparing to the SA results we have used
the Kendall $\tau$-c \cite{Kendall_c} correlation coefficient while we have used the
Kendall $\tau$-a \cite{Kendall_a} correlation coefficient in all the other cases.}
\label{table2}
\end{table}
In Figure $\ref{figure5}$a, we display $\Delta s$ for each node of the
American Airlines-United Airlines duplex network. The score seems informative.
If the top-ranked nodes are damaged deterministically (see Figure
$\ref{figure5}$b), the distribution $\pi(R)$ of the size of the giant
component becomes more peaked around the 
value $R=1/N$. 
If the top-ranked nodes are safeguarded (see Figure $\ref{figure5}$c),
the robustness of the entire system
in greatly improved. 
In fact, safeguarding the top-ranked node (Chicago O'Hare Airport, ORD) only
is already sufficient to observe a unimodal distribution $\pi(R)$
with peak at $R=0.3>R_c$.

In order to investigate whether the top-ranked nodes according to
$\Delta s$ have some relation with the nodes identified in solutions
to the optimal percolation problem, 
we performed systematic comparisons between the 
top  5 nodes according to $\Delta s$ and various methods 
used in optimal percolation~\cite{Radicchi2017}. 
We find that the top-ranked nodes correspond with good accuracy to the
nodes in the optimal structural 
set detected by Simulated Annealing optimization~\cite{Radicchi2017}. High correlation
(measured using Kendall $\tau$) is
also found with sets determined using other state-of-the-art
techniques (see  Table $\ref{table2}$). 
These  include the High Degree (HD) and the High Degree Adaptive (HDA) algorithms based on the product (HDp,HDAp) or the sum (HDs,HDAs) of the node degree in the two layers, and   the duplex network version of the Collective Influence (CI) algorithm  based on the product (CI$\ell$p) or on the sum (CI$\ell$s) of the CI scores of single layers (see Supplementary Information for details).     
   In the Supplementary Information,  we present the same
   type of analysis for
   for the United Airlines-Delta Airlines duplex
   network yielding similar conclusions.

\section{Conclusions}

We explored the large deviation properties of percolation of real 
finite multiplex  networks.  This approach 
 {consists in looking at the distribution $\pi(R)$ of the size $R$
of the MCGC as a function of the probability $p$ 
that a node is not initially damaged. 
The motivation of the study
finds its roots from the obvious inability of the mean value $\bar{R}$ to capture large
fluctuations that arise in finite-size systems.
Although the use of $\pi(R)$ is required to fully characterize the
properties of the percolation transition in real multiplex networks, 
in the paper, we demonstrated that most of the
system robustness can be 
understood by combining the information provided already by $\bar{R}$
with the complementary metric $\hat{R}$, i.e., the mode of the
distribution $\pi(R)$. $\hat{R}$} 
reveals the intrinsic fragility of a real multiplex 
displaying a discontinuity as a function of the
probability $p$ 
that a node is not initially damaged. This discontinuity characterizes
the position of an effective 
critical point $p=p_c$ where the distribution $\pi(R)$ of the sizes
$R$ of the MCGC is bimodal and displays two local maxima of the same
height at $R=1/N$ (indicating that the network is totally dismantled)
and at $R=R_c\gg 1/N$ (indicating that the network has a significantly
large MCGC). Therefore, for $p=p_c$, the possible outcome of an
initial damage is very uncertain.

The large deviation approach to percolation allows us  to characterize
the correlations among the state 
of different nodes in the network and the fluctuations 
in the state of single nodes measured by the so called specific heat
of percolation. 
Note that here we indicate by  state of a node its inclusion or
exclusion from the MCGC  resulting from   
a given realization of the initial damage. We show that nearest 
neighbor nodes display an average correlation that has a maximum for
a value of $p$ close to 
the percolation threshold  $p_c$. 

Finally, we focused our attention on the destiny of the MCGC at
$p=p_c$ 
proposing an algorithm able to detect some special nodes. The
safeguard of these nodes can ensure with high probability that the
most likely outcome is 
$\hat{R}>R_c$ and that the 
total dismantling of the network has a suppressed probability.
The proposed algorithm was tested on real datasets showing the
efficiency 
of the proposed safeguarding procedure. We further showed the set of top-scoring
nodes is almost identical to those found as solutions to the optimal
percolation problem.

\section*{Acknowledgements}
F.R. acknowledges support from the National Science Foundation (CMMI-1552487), and 
from the US Army Research Office (W911NF-16-1-0104).

\renewcommand{\theequation}{SM\arabic{equation}}
\setcounter{equation}{0}
\renewcommand{\thefigure}{SM\arabic{figure}}
\setcounter{figure}{0}
\renewcommand{\thetable}{SM\arabic{table}}
\setcounter{table}{0}
\renewcommand{\thesection}{SM\arabic{section}}
\setcounter{section}{0}

\section*{SUPPLEMENTARY INFORMATION}

\subsection*{Further investigation of the robustness of the American Airlines-United Airlines duplex network}

In this section we provide  some additional detail that contributes to the establishment of the robustness of the American Airlines-United Airlines duplex network (data from \cite{Radicchi2015}). However there is nothing specific about this dataset and the analysis that we outline here can be equivalently performed on any other duplex network.

\subsubsection*{Fluctuations around $\bar{R}$ and $\hat{R}$}
We have shown in the main text   (see Figure 2)  that  the typical size $\hat{R}$ of the MCGC  reveals the intrinsic fragility of the American Airlines-United Airlines duplex network. In fact it displays a clear discontinuity at $p=p_c$ while the average size $\bar{R}$ of the MCGC has a smooth profile. However we have also shown in  Figure 2 (main text)  that for $p=p_c$ the outcome of an initial damage is very unpredictable since the distribution $\pi(R)$ is bimodal displaying two maxima at $R=1/N\simeq 0.014$ and $R=R_c=0.27\gg 1/N$ with 
 \bea
 \pi(R=1/N)=\pi(R=R_c).
 \eea
 Here we estimate the expected fluctuations by measuring as a function of $p$ the standard deviation $\sigma_{\bar{R}}$ around the average size of the MCGC $\bar{R}$ and  the square-root deviation $\sigma_{\hat{R}}$ around the typical size $\hat{R}$ of the MCGC, i.e.  
\bea
\sigma_{\bar{R}}&=&\sqrt{\sum_{R}(R-\bar{R})^2 \pi(R)}, \\
\sigma_{\hat{R}}&=&\sqrt{\sum_{R}(R-\hat{R})^2 \pi(R)}.
\eea
\begin{figure*}[!htb]
\centering
\includegraphics[width=0.8\linewidth]{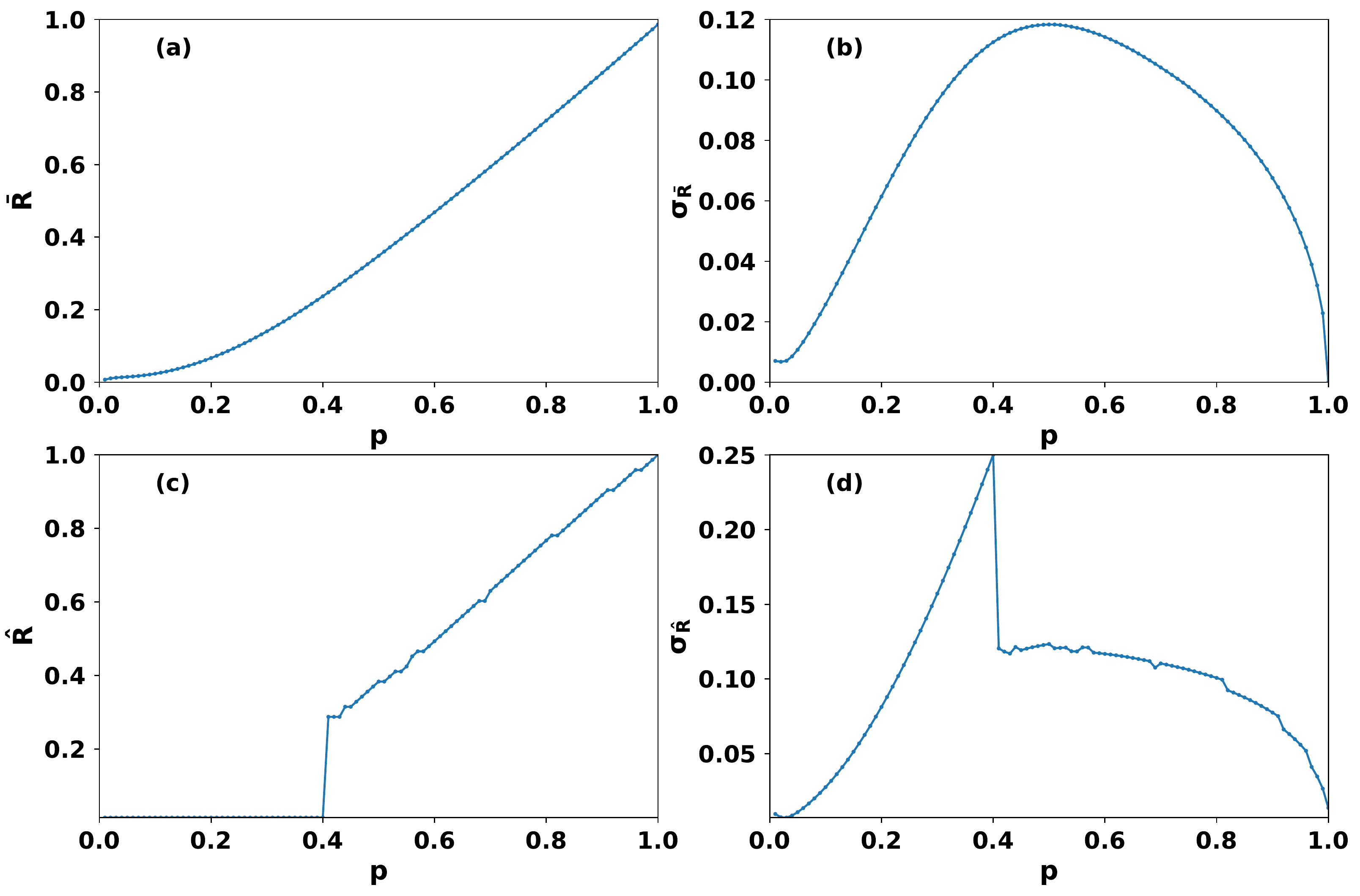}
\caption{The mean value of the MCGC size $\bar{R}$ (panel a) and average value $\bar{R}$ (panel b) of the American Airlines-United Airlines duplex network are plotted  as a function of $p$.  The square root  fluctuations $\sigma_{\hat{R}}$ (panel c)  and $\sigma_{\bar{R}}$ (panel d) of the size of the MCGC $R$ respect to $\bar{R}$  and $\hat{R}$  are plotted versus $p$. These results have been obtained by performing $Q=10^6$ realizations of the initial damage.}
\label{figureS1}
\end{figure*}

In Figure $\ref{figureS1}$ we display these quantities as a function of $p$ together with $\bar{R}$ and $\hat{R}$.
It is to be noted that $\sigma_{\hat{R}}$ has a jump at $p=p_c$ and reveals that for $p<p_c$ large fluctuations of the size of the MCGC can be observed.

\subsubsection*{How likely is the maximum likely outcome?}

\begin{figure}[!htb]
\centering
\includegraphics[width=0.45\linewidth]{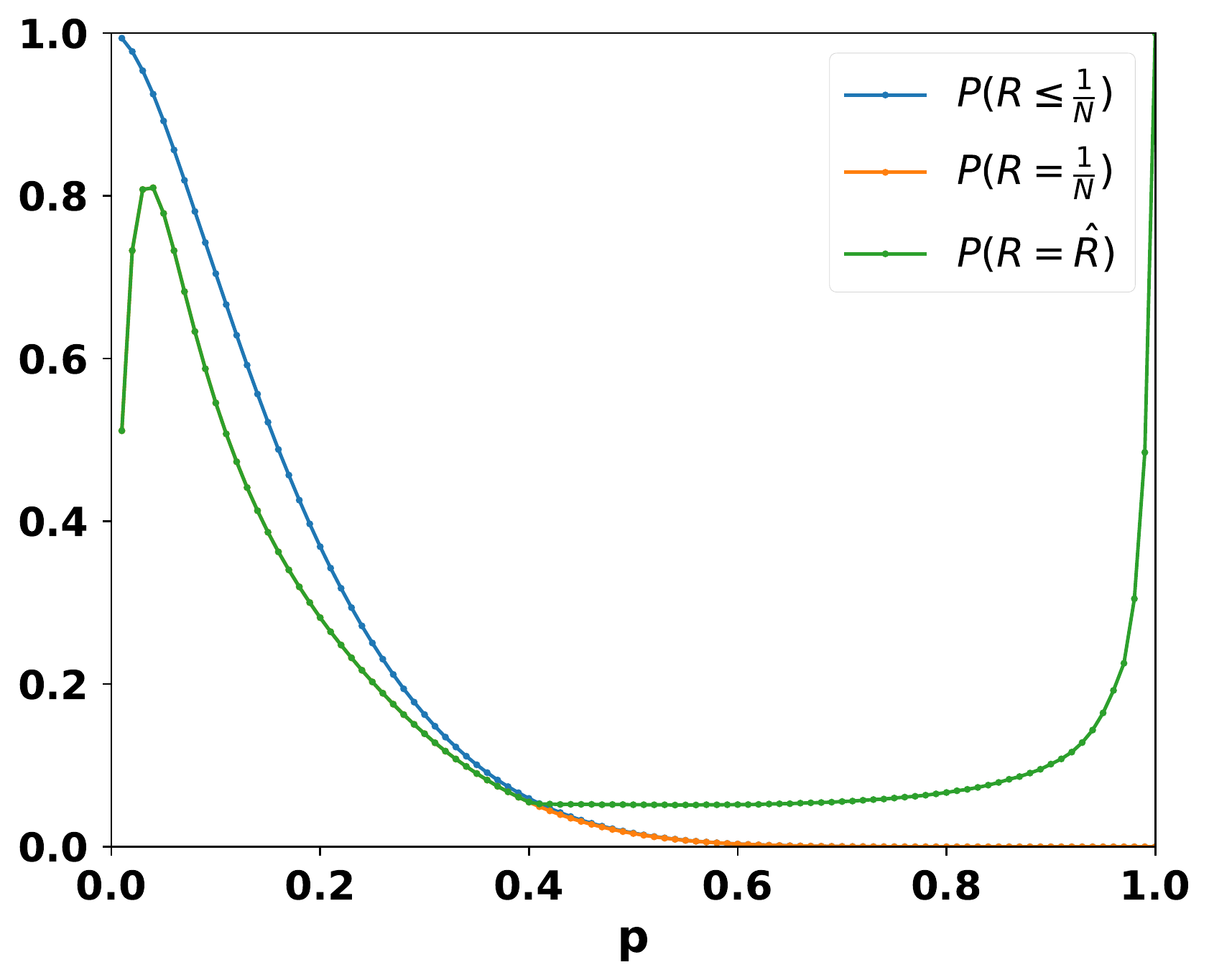}
\caption{The probabilities $P(R=1/N)$, $P(R\leq 1/N)$ and $P(R=\hat{R})$ of the American Airlines-United Airlines duplex network are plotted as a function of $p$. The probability that the network is totally dismantled $P(R\leq 1/N)$ is a monotonically decreasing function of $p$. For $p\leq 0.4$ the most likely outcome $\hat{R}=1/N$, therefore $P(R=\hat{R})=R(R=1/N)$.  As $p$ increases above $p=p_c$ the probability $P(R=\hat{R})$ of the most likely outcome $R=\hat{R}$ is at first not dependent on the value of $p$ while subsequently for values of $p$ approaching one it increases significantly. These results are obtained by performing $Q=10^6$ realizations of the initial damage.}
\label{figureS2}
\end{figure}

The large deviation study of percolation consists in analyzing the full probability distribution  $\pi(R)$ that the MCGC has size $R$ after an inflicted damage occurs on each node with probability $f=1-p$.
However in a number of cases it is useful to extract from $\pi(R)$ some statistical information that can synthetically indicate major aspects related to the robustness of the duplex network under study.
To this end here we consider the
probability 
$$P(R=\hat{R})= \pi(R=\hat{R}) / N$$ 
of the most likely size of the MCGC and we compare this quantity with
the probability 
$$P(R=1/N)= \pi(R=1/N) / N$$ 
that the MCGC is formed by only a single node (see Figure $\ref{figureS2})$.
Note that an initial damage of the nodes completely dismantles a duplex network if the size of the MCGC is $R=1/N$, nevertheless also MCGC of size $R=0$ can be observed if the initial damage is so severe that all the nodes of the network are initially damaged. Since MCGC of size $R=0$ are actually occurring with high probability for very small $p$    we have also considered the probability $P(R\leq 1/N)=P(R=1/N)+P(R=0)$ (see Figure $\ref{figureS2}$).
We observe that $P(R=\hat{R})=1$  for $p=1$ and decays rapidly as $p$ decreases  reaching a plateau persistent up to $p=p_c$. At $p=p_c$ the most likely outcome becomes $\hat{R}=1/N$ corresponding to a complete dismantling of the network. The probability that the MCGC is dismantled and $R\leq 1/N$ is monotonically increasing as $p$ approaches zero.

\begin{figure*}
\centering
\includegraphics[width=0.98\linewidth]{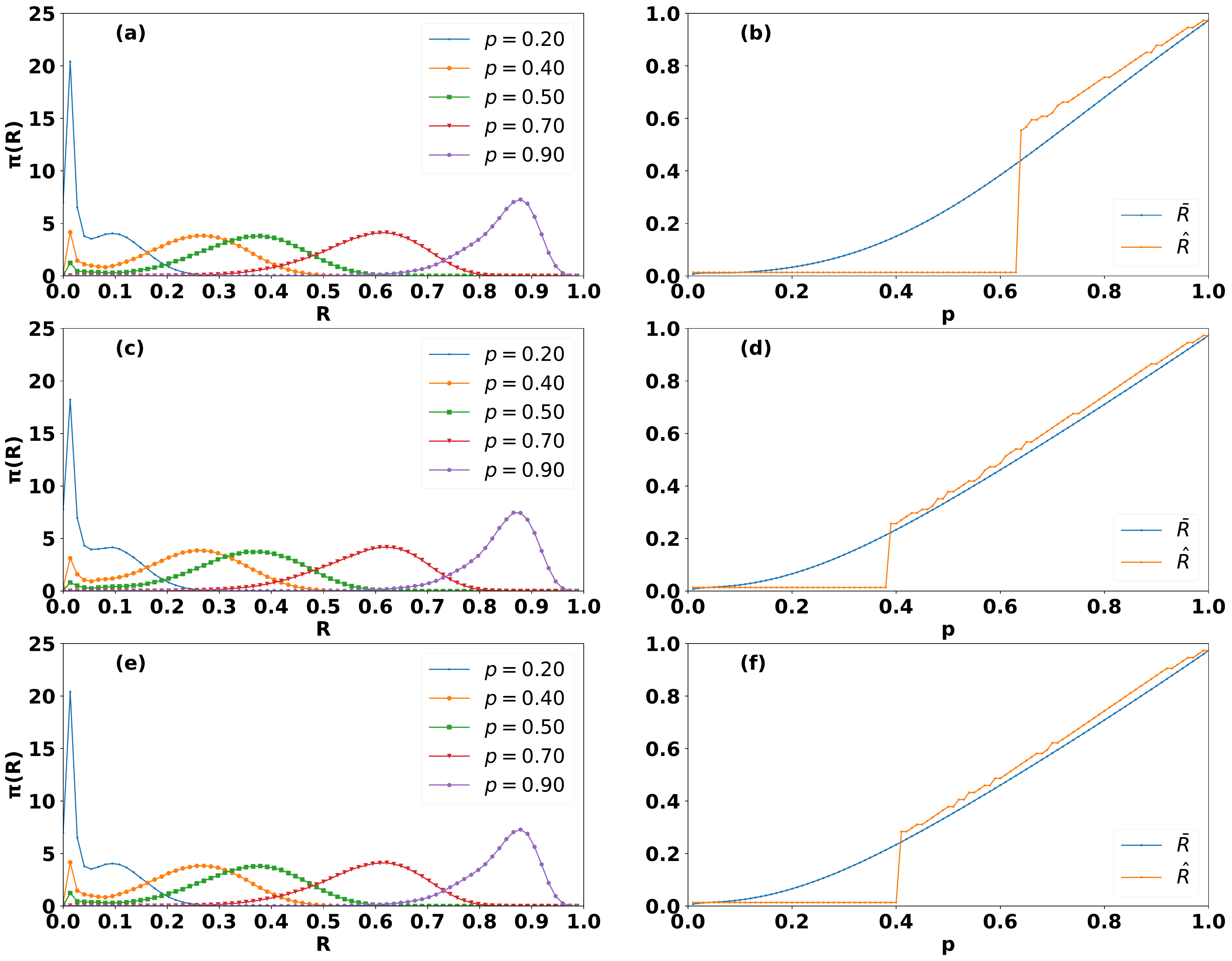}
\caption{{ Distribution $\pi(R)$,$\bar{R}$ and $\hat{R}$ for different null models of the American Airlines-United Airlines duplex network.} The distribution $\pi(R)$ of the size of the MCGC $R$ and the average ($\bar{R}$) and typical size ($\hat{R}$)  of the MCGC are plotted for  the null models: (1) randomization of the replica nodes (panels a,b), (2) independent randomization of each layer (panels c,d) (3) randomization preserving multidegree sequence (panels e,f) described in the text.These results have been obtained by performing $Q=10^6$ realizations of the initial damage.}\label{figure_nullmodels}
\end{figure*}
\subsubsection*{Distribution $\pi(R)$ for null models of the American Airlines-United Airlines multiplex network}
\label{secrand}

In order to investigate whether the bimodality of the distribution $\pi(R)$ is affected by structural correlations of the multiplex networks we have considered three types of possible randomization procedures \cite{Bianconi_book}.
\begin{itemize}
\item[(1)]{\it Randomization of the replica nodes}
We have kept the same networks but we have reshuffled the labels of the nodes in the second layer, removing inter-layer degree correlations.
\item[(2)] {\it Independent randomization of each layer}
We have randomized each layer independently keeping the same degree sequence. In this way we keep the same inter-layer degree correlations and we remove intra-layer degree correlations and link overlap.
\item[(3)]{\it Randomization preserving multidegree sequence}
We have randomized the multiplex network keeping the same multidegree sequence \cite{PRE,Bianconi_book}. In this way we keep the same level of link overlap and inter-layer degree correlations.
\end{itemize}
We have applied the three randomization procedures to the American Airlines-United Airlines multiplex network.
In Figure \ref{figure_nullmodels} we plot the distribution $\pi(R)$,$\bar{R}$ and $\hat{R}$ measured over $Q=10^6$ realizations of the initial damage for a single instance of each randomized null model.
We observe always a bimodality of the $\pi(R)$ distribution and we observe that the randomization that keeps the same level of overlap and intra-layer degree correlations best approximate the real results.

\begin{figure}[!htb]
\centering
\includegraphics[width=0.9\linewidth]{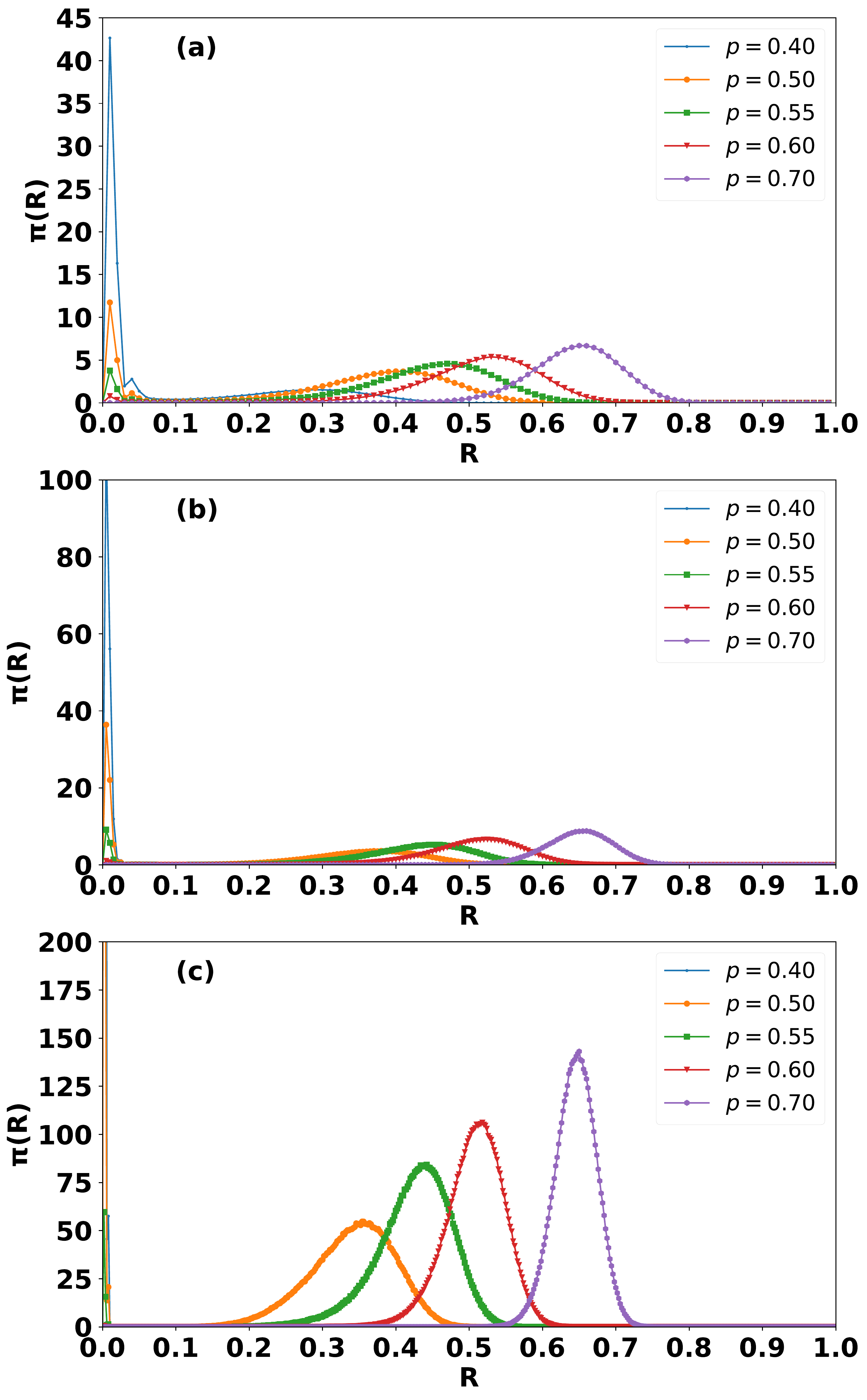}
\caption{The distribution $\pi(R)$ of the size of the MCGC $R$ is plotted for different value of $p$ for Posson multiplex networks with average degree $z=5$. The different panels refer to different network size $N$: $N=100$ (panel a) $N=200$ (panel b) and $N=500$ (panel c).These results have been obtained by performing $Q=10^6$ realizations of the initial damage.}
\label{Poisson_distr}
\end{figure}
\subsection*{Distribution $\pi(R)$ as a function of the network size on Poisson multiplex networks}

In order to explore the dependence of the distribution $\pi(R)$ as a function of the network size we have considered Poisson networks with different number of nodes $N$ and average degree $z=5$.
As noted in the main text, when the size $N$ of the network increases the interval of values of $p$ for which we observe a significant difference between the average size  $\bar{R}$  and the typical size $\hat{R}$ of 
the MCGC is reducing.  Here we note that additionally, as the network size increases the distribution $\pi(R)$ becomes bimodal for a narrowing range of values of $p$ (see Figure \ref{Poisson_distr}). Moreover, as $N$ increases the bimodal distribution $\pi(R)$ develop two well separated modes (see Figure \ref{Poisson_distr}).
 
\subsection*{State of the art algorithms for the Optimal Structural Node Set}

In this section we describe a number of state of the art algorithms \cite{Radicchi2017,Mendes2018} to detect the optimal structural node set or to rank the nodes according to their likelihood to be found in the optimal structural node set.
These algorithms include the duplex network version of the algorithms: Simulated Annealing (SA), High Degree (HD), High Degree Adaptive (HDA) and Collective Influence (Cl). 
While the SA algorithms provide a reasonable tight upper bound ot the optimal structural nodes set, the computational time necessary to achieve good results is significant. Therefore the other alternative algorithms that provides more greedy ways to rank the nodes in the optimal structural node set are also of relevant practical use  since they are significantly faster.

 \subsubsection*{High Degree (HD)}

On single networks the High Degree (HD) algorithms rank the nodes according to their degree.
On a duplex network the High Degree (HD) algorithm is typically modified in two different ways \cite{Radicchi2017}. The first algorithm (HDp) ranks the nodes by assigning to each node a score equal to the  the product of the degrees of the nodes in the two layers. In the second algorithms (HDs)  ranks the nodes by assigning to each node a score  equal to  the sum of its degrees in the two layers.

\subsubsection*{High Degree Adaptive (HDA)}

The High Degree Adaptive (HDA) algorithm is usually presented as an improvement of the High Degree (HD) algorithm. According to the HDA algorithm at each time step $t$ the node  $i$  with the highest HD score is associated to a rank $r_i=t$ and included in the structural node set, i.e. the node is  damaged an all its links are removed from the network.  Subsequently  the HD scores are  recomputed among the non damaged nodes of the network  until the network is completely dismantled.  In this work we have considered two different versions of the HDA algorithm proposed in Ref. \cite{Radicchi2017}: in the first one (HDAp) the score of each node is given by  the  product of its degrees in the two layers,  in the second one (HDAs) the score of each node is given by the the sum of the degrees in the two layers.

\subsubsection*{Collective Influence (CI)}

The CI algorithm \cite{Makse} on a single network assign to each node $i$ of the network a score given by 
\be 
{ Cl}_i(\ell) = (k_i-1) \sum_{j\in{\mathcal N}_{\ell}(i)} (k_j - 1) \; , 
\ee
where ${\mathcal N}_{\ell}(i)$ indicates the set of nodes at distance $\ell$  from $i$.
The CI algorithm is adaptive. This means that at each time $t$ the node with highest score is assigned a rank $r_i=t$, the node is included in a node structural set, i.e. it is damaged and all its links are damaged and finally the scores  are recalculated.
The algorithm ends when  the network is dismantled.

To adapt the CI algorithm to duplex networks we adopt the algorithms CI$\ell$p and CI$\ell$s proposed in \cite{Radicchi2017}. The  CI$\ell$p associates  to each node   the product of its CI scores in each  layer, in the CI$\ell$s  instead associates  to each node   the sum of its CI scores in each  layer,

Typically the CI algorithm on duplex networks prescribes to evaluate CI$\ell$p and CI$\ell$s scores
for different values of $\ell$ and consider the minimal structural set defined by the different  versions of the  algorithm.

However  given the small diameter of  the real duplex network taken under consideration in this paper here it makes no practical sense to extend this analysis beyond the distance $\ell=1$.
 
\subsubsection*{Simulated Annealing (SA)}

A Simulated Annealing algorithm (SA) can be used \cite{Radicchi2017,Mendes2018} for identifying a  structural node set that constitute a reasonably tight upper bound to the optimal (i.e. minimal) structural node set.
To this end we have used the SA algorithm proposed in Ref. \cite{Radicchi2017}.
To each node we associate the variable $s_i=0$ if the node is initially damaged and $s_i=1$ if the node is not initially damaged. The SA algorithm  detect the structural node set by  minimizing  the energy 
\bea
E=\gamma \sum_{i=1}^N(1-s_i)-\frac{R}{N}
\eea
with the parameter $\gamma$ measuring the  cost of removing a node from the multiplex network set like in Ref. \cite{Radicchi2017} to $\gamma=0.6$.
For the details used in the SA protocol we refer the reader to the Supplementary Information of Ref. \cite{Radicchi2017}.

\subsection*{Safeguarding of the MCGC in the United Airlines-Delta Airlines duplex network}

In this section we apply the ranking of the nodes according to $\Delta s$ in order to detect the nodes that are more relevant for determining the size of the MCGC of the United Airlines-Delta Airlines duplex network (data from \cite{Radicchi2015}).

\begin{figure*}[!htb]
\centering
\includegraphics[width=0.98\linewidth]{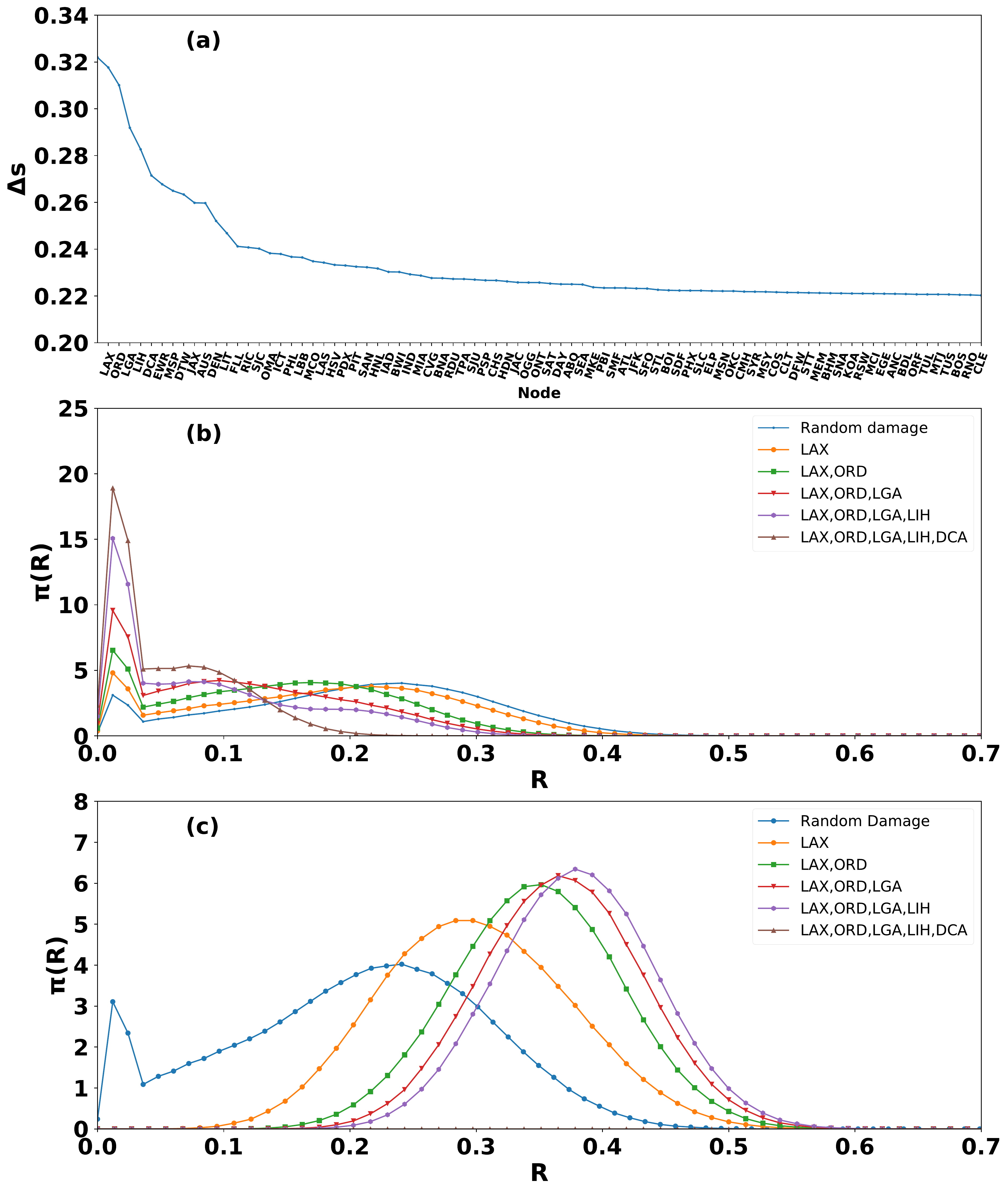}
\caption{{Effect of the damage and the safeguard of the top-ranked nodes on the robustness of the American Airlines-Delta Airlines duplex network.} The centrality measure $\Delta s$ for each of the $N=84$ airports of the American Airlines-Delta Airlines dataset is shown in panel (a). The centrality measures is evaluated by considering $Q=10^6$ realization of the initial damage at $p=p_c=0.37$ taking  $R^{\star}=1/\sqrt{N}<R_c=0.25$.   
The distribution $\pi(R)$ of the size $R$  of the MCGC at $p=p_c$ is compared to the distribution $\pi(R)$ obtained when the top-ranked nodes according to $\Delta s$ are damaged for sure (panel (b)) or safeguarded (panel (c)) while the other nodes are damaged with probability $p=p_c=0.37$. The distribution in panels (b) and (c) are obtained considering $Q=10^6$ realizations of the initial damage.}\label{figureS5}
\end{figure*}

Figure $\ref{figureS5}$a display the value of $\Delta s$ for each airport in the duplex network while the other panels of the same figure   display the  distribution $\pi(R)$  when the nodes with higher score $\Delta s$ are subsequently removed (see Figure  $\ref{figureS5}$b) or subsequently safeguarded (see Figure $\ref{figureS5}$c).

Table $\ref{tableS1}$ compares the ranking of the top airports ranked according to $\Delta s$ with the results obtained with the SA, HDp, HDs, HDAp, HDAs, CI1p, Cl1s. From this table it is clear that also in this case the nodes with high $\Delta s$ are largely correlated with the nodes in the optimal structural node set.

\begin{table}[!htb] \centering
\ra{1.6}
\begin{tabular}{@{}|ccccccccc|@{}}\toprule
\hlineB{3}
Airport & $\Delta s$ & SA & HDp & HDs & HDAp & HDAs & CI1p & CI1s \\
\hline
LAX & 0.3220 & 0 &1 &2 & 1 & 2 & 2 & 4 \\
ORD & 0.3176 & 0 & 6 &5 & 6 & 5& 17 & 10 \\
LGA & 0.3101 & 0 & 5 & 9 & 5 & 8 & 10 &11 \\
LIH & 0.2919 & 0 & 3 & 6 & 3 & 6 & 1 & 5 \\
DCA & 0.2827 & 0 & 21 & 8 & 8 & 9 & 39 & 23 \\
\hline
Kendall $\tau$ &1 & 1 &0.40	& 0.60	&0.40 &0.80 & 0.20& 0.60\\  
\hlineB{3}
\bottomrule
\end{tabular}
\caption{The top 5 airports in the United Airlines-Delta Airlines duplex network according to the centrality measure $\Delta s$ are listed together with their corresponding classification $\{s_i^{\star}\}$ according to the SA algorithm ($s_i^{\star}=0$/$s_i^{\star}=1$ indicating that node $i$ belongs/not belongs to the optimal node set) and their rank according to the HDp, HDs, HDAp, HDAs, CI1p and  CI1s algorithms.The last row indicates the Kendall-$\tau$ correlations among the ranking of these 5 airports according to $\Delta s$ and each of the other state of the art algorithms. Note that when comparing to the SA results we have used th Kendall $\tau$-c \cite{Kendall_c} correlation coefficient while we have used the Kendall $\tau$-a \cite{Kendall_a} correlation coefficient in all the other cases.}
\label{tableS1}
\end{table}

\begin{thebibliography}{99}

\bibitem{Bianconi_book}
G. Bianconi, {\it Multilayer Networks: Structure and Function} (Oxford University Press, Oxford, 2018).

\bibitem{PhysRep}
S. Boccaletti  et al., The structure and dynamics of multilayer networks. {\it Physics Reports} {\bf 544}, 1 (2014).

\bibitem{Kivela}
M. Kivel\"a, A. Arenas, M. Barthelemy, J. P. Gleeson, Y. Moreno and M.A.  Porter,  Multilayer networks. {\it Jour. Comp. Net.} {\bf 2}, 203 (2014).
\bibitem{Goh_review}
K. M. Lee, B. Min and K.I.  Goh,  Towards real-world complexity: an introduction to multiplex networks.  {\it Eur. Phys. Jour. B} {\bf 88}, 48 (2015).
\bibitem{PRE}
G.  Bianconi,   Statistical mechanics of multiplex networks: Entropy and overlap.
{\it Phys. Rev. E} {\bf 87}, 062806 (2013).

\bibitem{Havlin}
S. V. Buldyrev, R. Parshani, G. Paul, H. E.  Stanley and S.  Havlin,  Catastrophic cascade of failures in interdependent networks. {\it Nature} {\bf 464}, 1025 (2010).



\bibitem{Dorogovtsev}
G. J. Baxter, S. N.  Dorogovtsev, A. V.  Goltsev adn J. F. F.  Mendes,   Avalanche collapse of interdependent networks. {\it Phys. Rev. Lett.} {\bf 109}, 248701 (2012).

\bibitem{Goh}
B.  Min, S. D.   Yi, K.-M.  Lee and K.-I.  Goh, 
Network robustness of multiplex networks with interlayer degree correlations.
{\it Phys. Rev. E} {\bf 89}, 042811 (2014).


\bibitem{Son}
S.-W. Son, G.  Bizhani, C.  Christensen, P. Grassberger, M.  Paczuski,  
Percolation theory on interdependent networks based on epidemic spreading.  
{\it EPL (Europhysics Letters)} {\bf 97}, 16006 (2012).

\bibitem{HavlinEPL}
R. Parshani, C.  Rozenblat, D. Ietri, C. Ducruet and S. Havlin, 
 Inter-similarity between coupled networks. 
{\it EPL (Europhysics Letters)} \textbf{92}, 68002 (2010).

\bibitem{Havlin2}
R. Parshani, S. V. Buldyrev and S.   Havlin, 
 Interdependent networks: Reducing the coupling strength leads to a change from a first to second order percolation transition. {\it Phys. Rev. Lett.} {\bf 105}, 048701 (2010).


\bibitem{BD1}
G. Bianconi, S. N.  Dorogovtsev and J. F. F.  Mendes, 
Mutually connected component of networks of networks with replica nodes.
{\it Phys. Rev. E} {\bf 91}, 012804 (2015).


\bibitem{BD2}
G. Bianconi and S. N.    Dorogovtsev,   
 Multiple percolation transitions in a configuration model of a network of networks. 
{\it Phys. Rev. E} {\bf 89}, 062814 (2014).
\bibitem{RadicchiBianconi}
F. Radicchi and  G.  Bianconi,  Redundant interdependencies boost the robustness of multiplex networks. {\it Phys. Rev. X} {\bf 7}, 011013 (2017).

\bibitem{BianconiRadicchi}
G. Bianconi and F.   Radicchi,  Percolation in real multiplex networks. {\it Phys. Rev. E} {\bf 94}, 060301 (2016).



\bibitem{Cellai2016}
D. Cellai, S. N. Dorogovtsev and G.  Bianconi, 
Message passing theory for percolation models on multiplex networks with link overlap.
{\it  	Phys. Rev. E} {\bf 94}, 032301 (2016).
\bibitem{Baxter2016}
G. J.  Baxter, G.  Bianconi, R. A. da Costa, S. N.  Dorogovtsev and J. F. F.  Mendes, 
Correlated link overlaps in Multiplex Networks. 
{\it Phys. Rev. E} {\bf 94}, 012303 (2016).

\bibitem{Cellai2013}
D. Cellai, E. L\'opez, J. Zhou,  J. P. Gleeson and G.  Bianconi, 
Percolation in multiplex networks with overlap. 
{\it Phys. Rev. E} {\bf 88},  052811 (2013). 

\bibitem{Radicchi2015}
F. Radicchi, 
Percolation in real interdependent networks.
{\it Nature Phys.} {\bf 11}, 597 (2015).

\bibitem{Goh_comment}
B.  Min, S. Lee, K.-M. Lee and K.I.  Goh, 
Link overlap, viability, and mutual percolation in multiplex networks.
{\it Chaos, Solitons  Fractals} {\bf 72} 49 (2015).
\bibitem{Makse_NP}
S. D. S. Reis, Y. Hu, A. Babino, J. S. Andrade Jr., S. Canals, M. Sigman and H. A.  Makse, 
 Avoiding catastrophic failure in correlated networks of networks. 
{\it Nature Phys.} {\bf 10}, 762 (2014). 







 \bibitem{Souza}
 R. D'Souza and J.     Nagler,   
Anomalous critical and supercritical phenomena in explosive percolation.
{\it Nature Physics} {\bf 11}, 531 (2015).
\bibitem{Arenas}
J.  G\'omez-Garde\~nes, S. G\'omez, A. Arenas and Y.  Moreno, 
Explosive synchronization transitions in scale-free networks. 
{Phys. Rev. Lett.} {\bf 106},  128701 (2011).
\bibitem{Vito}
V. Nicosia, P. S. Skardal, A. Arenas and V.  Latora, 
Collective phenomena emerging from the interactions between dynamical processes in multiplex networks.
{\it Phys. Rev. Lett.} {\bf 118}, 138302 (2017).

\bibitem{RadicchiArenas}
F. Radicchi and A.  Arenas,  
 Abrupt transition in the structural formation of
interconnected networks. 
{\it Nature Phys.} {\bf 9}, 717 (2013).
\bibitem{Doro_kcore}
S. N.  Dorogovtsev, A. V.  Goltsev and J. F. F.  Mendes, 
K-core organization of complex networks.
{\it Phys. Rev. Lett.} {\bf 96},  040601 (2006).
\bibitem{Rizzo}
G. Parisi and T.    Rizzo, 
k-core percolation in four dimensions. 
{\it Phys. Rev. E} {\bf 78}, p.022101 (2008).
\bibitem{crit} 
S. N. Dorogovtsev, A. V.  Goltsev and J. F. F.  Mendes, 
Critical phenomena in complex networks.
Rev. Mod. Phys. {\bf 80}, 1275 (2008).

\bibitem{Lenka}
B.  Karrer, M. E. J.  Newman and L.  Zdeborov\'a, 
Percolation on sparse networks.
{\it  Phys. Rev. Lett. }{\bf 113},  208702 (2014).
 \bibitem{Wei}
W. K.  Chai, V. Kyritsis, K. V.  Katsaros and G.   Pavlou, 
 Resilience of interdependent communication and power distribution networks against cascading failures. 
 In IFIP Networking Conference (IFIP Networking) and Workshops, 2016 (pp. 37-45). IEEE (2016).

\bibitem{Fluct}
G. Bianconi,   Fluctuations in percolation of sparse complex networks. {\it Phys. Rev. E} {\bf 96}, 012302 (2017).

 
\bibitem{Bianconi2018}
G. Bianconi,  Rare events and discontinuous percolation transitions. {\it Phys. Rev. E} {\bf 97}, 022314 (2018)




\bibitem{Makse}
F. Morone and H.A.  Makse,  Influence maximization in complex networks through optimal percolation. {\it Nature} {\bf 524}, 65 (2015).
\bibitem{Dismantling}
A. Braunstein, L. Dall’Asta, G. Semerjian and L.  Zdeborov\'a, L.  Network dismantling. {\it Proc. Nat. Aca. Sci.} {\bf 113}, 12368 (2016).
\bibitem{Kahng}
D. Lee,  S. Choi, M. Stippinger, J. Kert\'esz, and B. Kahng, Hybrid phase transition into an absorbing state: Percolation and avalanches.{\it  Phys. Rev. E} {\bf 93}, 042109 (2016).

\bibitem{Krapivsky}
M. Kitsak, A. A. Ganin, D. A. Eisenberg, P. L. Krapivsky, D. Krioukov, D. L. Alderson and  I. Linkov,  Stability of a giant connected component in a complex network. {\it Phys. Rev. E} {\bf 97}, 012309 (2018).
\bibitem{Radicchi2017}
S. Osat, A. Faqeeh and F.   Radicchi,  Optimal percolation on multiplex networks. {\it Nat. Comm.}  {\bf 8}, 1540 (2017).

\bibitem{Mendes2018}
G. J. Baxter, G. Tim\'ar and J. F. F.  Mendes,  Targeted Damage to Interdependent Networks. arXiv preprint arXiv:1802.03992 (2018).

 
\bibitem{Connectome}
B. L.  Chen, BD. H.   Hall and D. B.   Chklovskii,  Wiring optimization can relate neuronal structure and function" {\it Proc. Nat. Aca. Sci.} {\bf 103}, 4723 (2006)
\bibitem{Manlio}
M. De Domenico, M.A.  Porter and  A.   Arenas,  MuxViz: A Tool for Multilayer Analysis and Visualization of Networks. {\it Jour.  Comp. Net.}  {\bf 3},  159 (2015).
\bibitem{SI}
Supplementary Information available at 
\bibitem{Parisi_Mezard}
M. M\'ezard, G. Parisi, and M. Virasoro, {\it Spin glass theory and beyond: An Introduction to the Replica Method and Its Applications} (World Scientific Publishing Company 1987).
\bibitem{Kendall_c}
K. J. Berry,  J. E. Johnston, S. Zahran, P. W.  Mielke,  Stuart's tau measure of effect size for ordinal variables: Some methodological considerations. {\it Behavior Research Methods} {\bf 41}, 1144 (2009).
\bibitem{Kendall_a}
M. A. Kendall, New Measure of Rank Correlation. {\it Biometrika} {\bf 30}, 81  (1938).

\end{thebibliography}
\end{document}